\documentclass[twocolumn]{aastex63}

 \usepackage{hyperref}
\usepackage{amsmath}
 \usepackage{xfrac,xcolor,hyperref,ulem}
\usepackage{datetime2}
 \usepackage{makecell}

\usepackage{comment}

\graphicspath{{./}{Figures/}}

 \received{\today\ \DTMcurrenttime}

\shorttitle{Neutral Fraction of the Warm Ionized Medium}
 \shortauthors{Kulkarni et al.}

\begin{document}

\title{Probing the Neutral Fraction of the Warm Ionized Medium via
[NI]\,$\lambda$5200}

\correspondingauthor{S.\ R.\ Kulkarni}
 \email{srk@astro.caltech.edu}

\author[0000-0001-5390-8563]{S.\ R.\ Kulkarni}
 \affiliation{Owens Valley Radio Observatory 249-17, California
 Institute of Technology,
Pasadena, CA 91125, USA}

\author{S.\ Noll}
 \affiliation{German Space Operations Center (GSOC), Deutsches Zentrum f\"ur Luft- und Reaumfahrt (DLR), Oberpfaffenhofen, Germany.}
 \affiliation{Institut f\"ur Physik, Universit\"at Augsburg, Augsburg, Germany}
 \affiliation{Deutsches Fernerkundungsdatenzentrum, Deutsches Zentrum 
  f\"ur Luft- und Raumfahrt (DLR), Oberpfaffenhofen, Germany}
\email{st@noll-x.de}

\author{W.\ Kausch}
 \affiliation{Institut f\"ur Astro- und Teilchenphysik, Universit\"at Innsbruck,
 Austria}
\email{wolfgang.kausch@uibk.ac.at}

\author{Soumyadeep Bhattacharjee}
 \affiliation{Department of Astronomy, 249-17 California Institute of 
 Technology, Pasadena, CA 91125, USA}
 \email{sbhatta2@caltech.edu}
 
\begin{abstract}
 Most of the ionized mass in the Milky Way is in the Warm Ionized
 Medium (WIM) and not in the bright H~II regions.  The WIM is traced
 by dispersion measure and has been extensively studied in recombination
 lines (primarily, H$\alpha$) and optical nebular lines (primarily,
 S$^+$ and N$^+$). The observations can be well explained by a
 photo-ionized nebula with a low ionization parameter.  It is
 generally thought that the source of ionization (and heating) of
 the WIM is due to Lyman continuum leaking from H~II regions which
 are concentrated in the Galactic plane.  The rays of the diffuse
 Galactic Lyman-continuum radiation field incident on the Warm
 Neutral Medium (WNM) are absorbed, forming an ionized skin.  In
 nebulae with low-ionization parameter the transition from ionized
 gas to neutral gas is gradual, unlike the case for H~II regions
 with their sharp Str\"omgren spheres.  The transition region is
 warm enough to excite oxygen and nitrogen atoms to emit
 [OI]\,$\lambda\lambda$6300,\,6363 and [NI]\,$\lambda\lambda$5198,\,5200.
 \citet{dm94} recognized the value of [OI]\,6300 as a diagnostic
 of the fraction of the diffuse continuum that is absorbed by the
 WNM and therefore constrains the fraction of the diffuse Lyman
 continuum that escapes to the halo.  Unfortunately, observations
 of Galactic [OI]\,6300 have been severely stymied by bright
 [OI]\,6300 airglow emission.  [NI]\,$\lambda\lambda$5200,\,5198
 has been a historically less popular probe because this doublet is less luminous than the oxygen doublet.  However, we point out
 that the [NI] airglow is two orders of magnitude smaller than that
 of [OI]. Furthermore, even in the presence of comparable airglow,
 the WIM [NI] emission can be inferred using the doublet intensity
 ratio for which a medium-resolution spectrometer such as the Local
 Volume Mapper will suffice.  Separately, in extragalactic systems,
 we note that $\lambda$6300/$\lambda$5200 is a robust measure of
 the O/N abundance ratio.   \\
\end{abstract}

\section{Background \&\ Motivation}
 \label{sec:Background}

In our Galaxy, the Warm Ionized Medium (WIM) hosts most of the
ionized gas and occupies perhaps a quarter of the disk volume. This
phase is seen in other galaxies but is given the moniker of Diffuse
Ionized Gas (DIG). See \cite{hdb+09} for a review of both the WIM
and the DIG.

The WIM is now thought to be ionized and powered by Lyman continuum
radiation leaking from OB stars. The leakage is due to holes
(``champagne flow")  in the ISM surrounding star-forming complexes
or due to some H~II regions being ``density bounded\footnote{The
ionizing rays run out of matter to ionize. In this case, the spectrum
of the leakage radiation will be softer relative to the escape
through the holes}''.  The leakage fraction is estimated to be one
sixth of the ionizing output of Galactic OB stars \citep{R84}.  The
resulting diffuse Lyman continuum propagates into the interstellar
medium (ISM) and upon encountering diffuse atomic hydrogen -- which
is likely to be the Warm Neutral Medium (WNM) because it has a
filling factor larger than that of the Cold Neutral Medium (CNM)
-- will start ionizing it.

\cite{M86} recognized that the primary difference between the WIM
and classical H~II regions is that the former has fewer ionizing
photons per H atom. A consequence of this ``dilute H~II" model
(which, in current parlance, is a model with low ionization parameter)
is that the WIM, unlike H~II regions, is partially ionized. This
difference readily explains the higher strengths of the [NII] and
[SII] lines, the absence of highly ionized species, and the higher
temperature of the WIM relative to H~II regions.

Consider a ray of the diffuse Lyman continuum field that is propagating
away from the Galactic plane along the perpendicular direction.  If
the ray only encounters the Hot Ionized Medium (HIM), then it can
freely reach the halo region.  This is certainly the case for
Galactic ``chimneys'' and ``worms" \citep{khr92,H94}.  Now consider
the case where the ray primarily propagates into the WNM (hydrogen
number density, $n_{\rm H}$) which, after all, has a filling factor
comparable to that of the HIM.  In steady state, the photon flux
density of this ray is balanced by recombinations, $\alpha n_en_{\rm
H^+}L$ where $\alpha$ is the recombination coefficient, $n_e$
($n_{\rm H^+}$) is the number density of electrons (protons) and
$L$ is the length of the ionized column.

In this transition layer, the partially ionized gas, heated by
photoionized electrons, will be warm.  The ionization ratio of
oxygen, O$^+$\!/O$^0$, thanks to the near coincidence of the
ionization potential of oxygen (13.6181\,eV) with that of hydrogen
(13.5984\,eV), coupled with a large charge transfer cross section,
results in ${\rm O^+\!/O^0}\approx (8/9) {\rm H^+\!/H^0}$ (see
\S14.7 of \citealt{D11}).  The ionization potential of nitrogen is
14.53\,eV, which is approximately 1\,eV higher than that of hydrogen.
Thus, N$^0$\!/N$^+$ should also be a good proxy for H$^0$\!/H$^+$.

In the ISM, carbon, being ionized by the stellar FUV field, is
present in both the WIM and WNM. The dominant metals that matter
in the transition region are oxygen, nitrogen, and neon.  Ne$^0$
lacks optical nebular lines, leaving us with
[OI]\,$\lambda\lambda$6300,\,6363 and [NI]\,$\lambda\lambda$5198,\,5200
as possible diagnostics.  Because of the higher abundance, the
oxygen doublet is brighter than the nitrogen doublet.  Within the
doublets, the brighter lines are [OI]\,6300 and [NI]\,5200.

\cite{M86} computed the expected [OI]\,6300 and [NI]\,5200 emission
from the WIM.  \cite{R89}, using the Wisconsin H$\alpha$ Mapper
(WHAM; \citealt{T97,rth+98}), a Fabry-P\'erot spectrometer designed
for sensitive studies of the WIM, observed one line-of-sight at low
latitude, devoid of H~II regions, and found the photon intensity
of [OI]\,6300 relative to that of H$\alpha$ was $<0.02$ whereas the
model of \cite{M86} predicted this ratio to be $\approx 0.07$.
\cite{dm94} resolved this discrepancy by invoking WNM clouds with
column density less than $n_{\rm H}L$ in which case there is no
transition layer, which then results in a reduction of [OI] and
[NI] emission.  A consequence of this assumption is that the diffuse
Lyman continuum that is incident on such thin WNM will exit the WNM
cloud and escape to the halo.  \cite{dm94} concluded their seminal
paper by noting that the strength of the [OI]\,6300 line can serve
as a proxy for the fraction of the diffuse Lyman continuum consumed
by the WNM and therefore, indirectly, constrain the fraction of the
diffuse Galactic Lyman continuum that escapes from the disk to the
halo.  Interestingly, it is now standard practice to include the
variable thickness of the neutral cloud as a fundamental parameter
in modeling the WIM (e.g., \citealt{shr+00}).

Unfortunately,  [OI]\,6300 is among the brightest\footnote{To be
accurate: there are many bright molecular lines (OH and O$_2$).
The atomic lines of oxygen are bright and also variable, being excited by electrons
(ionosphere). The auroral [OI]\,5577 line is
also very bright, as are the resonance OI\,$\lambda$1304 and the
semi-forbidden OI]\,$\lambda$1356  lines (and clearly seen in HST
spectra).} airglow line in the visible band.  Spectrometers with
very high spectral resolution (ideally, $\approx 300,000$) are
needed to separate the faint WIM [OI] nebular line from the bright
airglow line.  WHAM, designed for H$\alpha$ studies, has the necessary
spectral resolution (30,000) to separate the WIM H$\alpha$ from
geocoronal H$\alpha$, although at the cost of a 1-degree beam. In
contrast, most modern Integral Field Unit (IFU) spectrographs lack
such high spectral resolution.  This situation motivated us to
explore [NI]\,5200 as a possible alternative.  This line is nearly
six times weaker than the [OI]\,6300 line, mainly due to the smaller
abundance of nitrogen relative to that of oxygen.  However, {\it
the [NI] airglow line is about a hundred times fainter than the
[OI] airglow line.} Separately, for some reason, after \cite{M86},
WIM modeling papers (e.g., \citealt{dm94,shr+00}) stopped modeling
the [NI] 5200 line.  The purpose of this paper is to thoroughly
investigate the practicality of using [NI]\,5200 for WIM studies.

The paper is organized as follows.   In \S\ref{sec:IonizationModels}
we review the currently accepted photo-ionization model developed
for the WIM.  In \S\ref{sec:N_Lines} we review the nebular and
auroral lines of N$^0$ in some detail since, as noted earlier,
unlike [OI], the N$^0$ doublet is not widely discussed in the WIM
literature.  The airglow observations of [NI] and [OI] are summarized
in \S\ref{sec:Airglow_NI} and \S\ref{sec:OI}, respectively.  In
\S\ref{sec:Detectability} we review the prospects for studying the
WIM in O$^0$ and N$^0$ with the current suite of wide-field
spectrographs.  We conclude in \S\ref{sec:SummaryConclusions}.

\section{The Dilute H~II region  model for the WIM}
 \label{sec:IonizationModels}

Two parameters are sufficient to characterize low-density\footnote{in
such nebulae, radiative decays dominate over collisional de-excitations}
photo-ionized plasma: the ``ionization parameter", $u=n_\gamma/n_{\rm
H}$, where $n_\gamma$ is the density of ionizing photons and $n_{\rm
H}$ is the density of H nuclei (\citealt{dn79}) and the effective
stellar temperature of the radiation field, $T_*$.  Normal H~II
regions are modeled by $u\gtrsim 10^{-2}$ while $u\lesssim 10^{-3}$
is invoked for the WIM.

We adopted a slab\footnote{Since the publication of the three-phase
McKee-Ostriker global model (which invoked spheres with CNM cores
and WNM envelopes) there has been considerable evidence that the
WNM and CNM are distributed in sheets and fibers (see, for example,
\citealt{msr23} for a recent review on Galactic H~I).} geometry to
model the WIM.  The Galactic diffuse Lyman continuum field propagates
into a slab and gradually decays, as ionizing photons are used to
balance recombinations.  We first discuss a simplified 1-D model
focused solely on the ionization of hydrogen (\S\ref{sec:SimplifiedModel}).
This model has the value of providing physical insight.  We then
present a numerical model obtained with CLOUDY, a state-of-the-art
modeling tool (\S\ref{sec:CLOUDY}).  The latter approach has the
advantage of modeling line strengths at high fidelity.

\subsection{Simplified Model}
  \label{sec:SimplifiedModel}
  
Let $I_0$ be the photon intensity of the diffuse Lyman continuum
radiation field that is incident on a thick WNM slab.  The ionization
parameter is then $u=(\pi I_0/c)/n_{\rm H}$.  For numerical modeling,
it helps to reexpress equations with dimensional-less quantities
and this results in a characteristic dimensional-less parameter:
$\xi=(\pi I_0\sigma_{\rm pi})/(\alpha n_{\rm H})$. Here, $\sigma_{\rm
pi}$ is the average photoelectric cross section and $\alpha$ is
the hydrogen recombination coefficient. The astute reader will
recognize that $\xi$ is the ratio of the recombination timescale to the
photo-ionization timescale. 

\begin{figure}[htp] 	
 \plotone{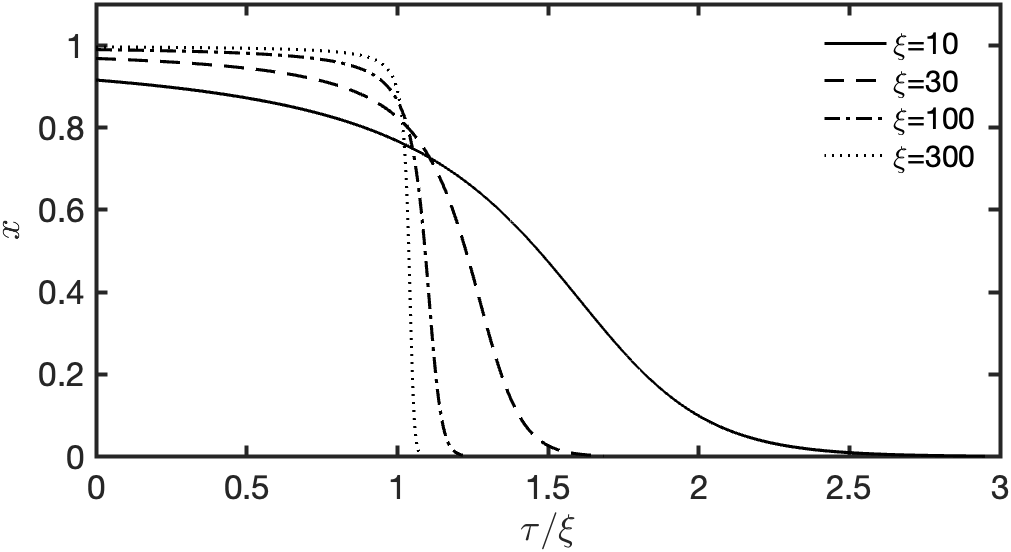}
  \caption{\small 
  Run of $x=n_{\rm H^+}/n_{\rm H}$, the ionization fraction, as a
  function of $\tau/\xi$. } 
 \label{fig:Slab_tau_x}
\end{figure}

The simplified model is fully presented in \S\ref{sec:1-D}.  Here,
we review the results.  For nominal parameters used in WIM modeling,
$\xi\approx 2\times 10^5u$.  The range of interest for WIM models
is $u=10^{-4}$ to $10^{-3}$, which corresponds to $\xi=20$ to 200.
The coordinate perpendicular to the slab is $z$ (``depth"). The
formal optical depth of the ionizing photons is $\tau=\sigma_{\rm
pi}n_{\rm H}z$.

In Figure~\ref{fig:Slab_tau_x} we present the profile of hydrogen
ionization.  For a plasma with a small $\xi$, the ionization fraction
at the top ($z=0$) surface, $x(0)$, is not close to unity.   Next,
the sharpness of the transition from an ionized to a neutral medium
depends on $\xi$. For small values of $\xi$, the transition is
gradual.

The simplified model has only one free parameter, $\xi$, whereas
CLOUDY has two free parameters, $u$ and $T_*$. CLOUDY can compute
ionization fractions of all species and temperature, whereas the
simplified model can compute only the ionization fraction of hydrogen.
However, on general grounds, we can conclude as follows:
 \begin{enumerate} 
    \item Given that the only source of
heating is photoionization, we can expect that a higher value of $u$
results in a higher gas temperature, $T$.
 \item The effective stellar
temperature of the radiation field, $T_*$, sets the typical energy
of the ionizing photons. Thus, for a fixed $u$, a higher $T_*$
results in a higher energy for a typical ionizing photon. This has
two effects. ({\it a}) The higher energy of the typical photon
results in a higher $T$.  ({\it b}) The photoelectric absorption cross
section decreases with increasing photon energy.  For a fixed value
of $\mathcal{I}_0$, the lower value of $\sigma_{\rm pi}$ results
in a smaller $\xi$.  Consider two nebulae both illuminated by the
same intensity, $I_0$, but $T_*$ for one is higher than that for
the other. The $\xi$ for the higher $T_*$ nebula will then be smaller
than that for the lower $T_*$.  Thus, in Figure~\ref{fig:Slab_tau_x},
we expect that the ionization profile of a nebula illuminated by
higher $T_*$ will be gradual relative to that of a nebula illuminated
by lower $T_*$ 
 \end{enumerate} 
In fact, as we will see in the next
subsection, these expectations are borne out by CLOUDY.

\begin{figure}[htbp]    
 \plotone{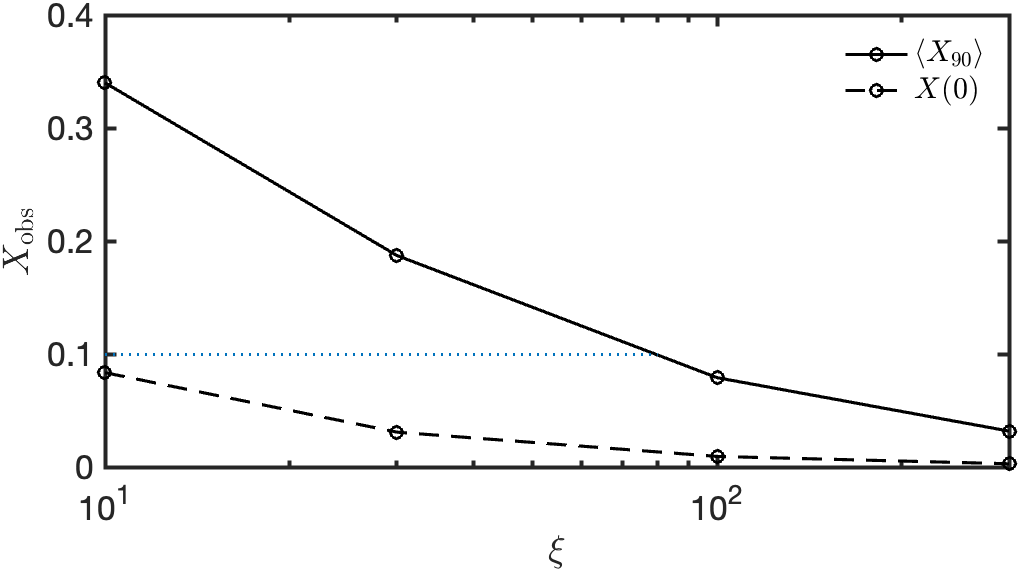}
  \caption{\small The run of $X(0)$ (the neutral fraction at the
  surface) and  $\langle X_{90}\rangle$ with $\xi$. Here, $\langle
  X_{90}\rangle$ is  the mean neutral fraction of the WIM over the
  neutral fraction range, $X=[X(0),0.9]$.  with $\xi$.  The dotted
  horizontal line corresponds to a hypothetical observed neutral
  fraction of 0.1. In this case, the observations constrain $\xi$
  to lie between 10 and 80.  }
 \label{fig:X0X90}
\end{figure}

The only heat input in this model is the kinetic energy of the
photo-ionized electrons. The dominant cooling is due to electrons
exciting ions.  As one proceeds into the slab, the electron density
goes down. Both heating and cooling decrease, and the electron
temperature remains roughly constant. However, at a low enough value
of $x$, the temperature drops and the line excitation rapidly drops.
In WIM modeling, $x=0.1$ is taken as the edge of the WIM layer (cf.\
\citealt{shr+00}). This boundary condition is usually marked by the
neutral fraction at the edge, $X_{\rm edge}$. Let $L$ be the depth
corresponding to $X_{\rm edge}=0.9$.  The lowest neutral fraction
is on the surface, $X(0)=1-x(0)$.  Let $\langle X_{90}\rangle$ be
the average neutral fraction between the top surface and the bottom
surface.  In Figure~\ref{fig:X0X90} we plot the run of $X(0)$ and
$\langle X_{90}\rangle$ with $\xi$.

Note that the plasma in the transition region is still warm (being
heated by the photoelectric process). As such, this region will
emit strongly in the [OI] lines.  However, as noted in
\S\ref{sec:Background}, \cite{R89} did not find a strong emission
of [OI] as predicted by the WIM model of \cite{M86}.  To solve this
problem of over prediction, \citet{dm94} proposed $X_{\rm edge}$ has
to be less than 0.9. By truncating the WIM layer [OI] emission can
be reduced to match the observations. So, to model the WIM, we now
need three physical parameters: $u$, $T_*$, and $X_{\rm edge}$.

For illustration, assume that from observations we have concluded
that the mean neutral fraction along a line-of-sight is $X_{\rm
obs}\approx 0.1$.  As can be gathered from Figure~\ref{fig:X0X90},
the allowed range is $10<\xi<80$.  If $\xi$ is 10 then the neutral
fraction on the top surface already has the inferred $X_{\rm obs}$.
The WIM slab cannot be thick since then $\langle X\rangle$ will
increase.  If $\xi\approx 80$, then the WIM slab must be thick
enough so that the bottom surface reaches $X=90\%$.  A thick slab
immersed in a field with $\xi>80$ will have a mean neutral fraction,
$\langle X_{90}\rangle<X_{\rm obs}$.

\subsection{Calculations with CLOUDY}
	\label{sec:CLOUDY}
	
The simplified model is adequate to give us an overall physical
understanding of the problem. However, it does not provide accurate
measures of the line intensity whose strength also depends on the
electron temperature (which is not captured by the simple model).
To this end, we employed CLOUDY \citep{fcg+17}.  

\begin{deluxetable}{lrr}[ht]
 \tablecaption{Cosmic Abundance}
  \label{tab:CosmicAbundance}
\tablewidth{0pt}
 \tablehead{
  \colhead{sp} &
  \colhead{Draine} &
  \colhead{Orion} }
\startdata
O  & 537 & 331 \\
N  & 74  & 33 \\
Ne & 93  & 46 \\
S  & 14  & 16 \\
\enddata
 \tablecomments{\small The abundance of species (O, N, Ne, S), relative
 to hydrogen,  in parts per million (ppm).
 Draine values are from Table~1.4 of \cite{D11}.
 The Orion values are from Table~1 of \cite{dm94}. }
\end{deluxetable}

As before, we assume a slab geometry.  The calculations were stopped
when the hydrogen ionization fraction reached $0.1$, corresponding
to $X_{\rm edge}=0.9$.  \cite{dm94} and \cite{shr+00} used the
cosmic abundance derived from the Orion nebula \citep{ptr92}. We
adopted the abundances from \cite{D11}. As can  be seen from
Table~\ref{tab:CosmicAbundance} there are significant differences
between these two lists.

\begin{figure}[htbp]  
 \plotone{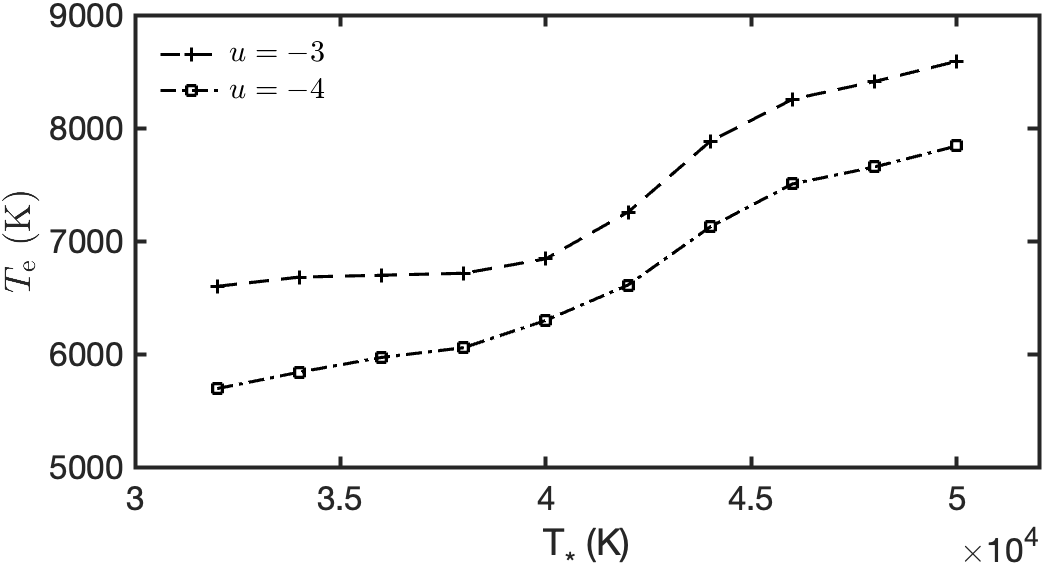}
  \caption{\small Run of electron temperature, $T$, with $T_*$. }
 \label{fig:Tstar_Tgas}
\end{figure}

\begin{figure}[htbp]    
 \plotone{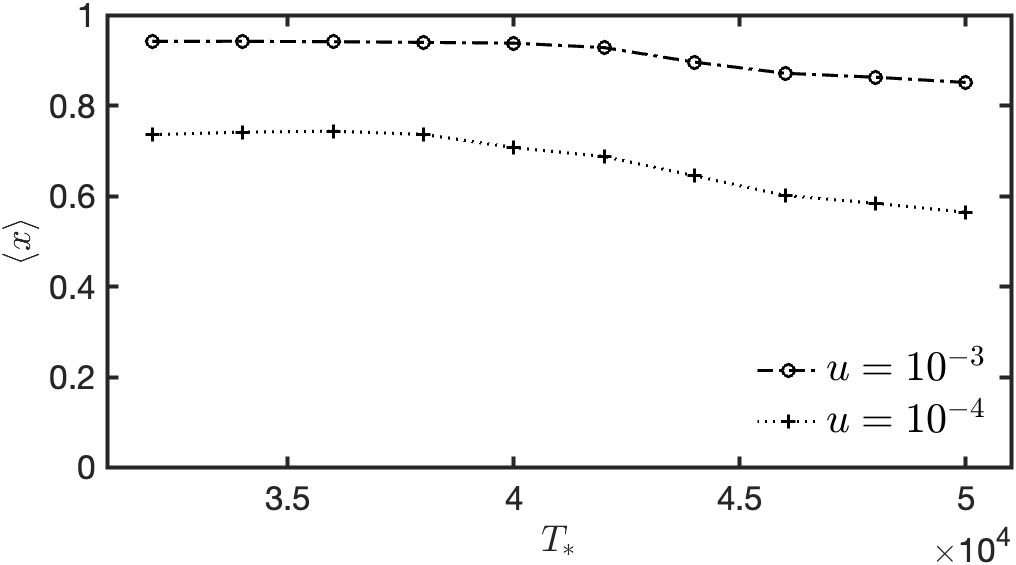} 
  \caption{\small Run of average ionization fraction, $\langle
  x\rangle=\int_0^L xdz/\int_0^L dz$, with $T_*$ for two values of
  $u$.  }
 \label{fig:x_avg} 
\end{figure}

We ran CLOUDY (version  C23; \citealt{cbc+23}) for $u=[10^{-3},10^{-4}]$
and $T_*=[3.2,...,5]\times 10^4$,K.  The fiducial parameters for
the WIM are $\log(u)=-4$ and $T_*\approx 3.8\times 10^4\,$K
\citep{shr+00}.  We find, as expected, that for a fixed value of
$u$, the temperature of the WIM scales with $T_*$
(Figure~\ref{fig:Tstar_Tgas}). This makes sense since the photon
energy per H atom increases with $T_*$.  Next, as can be seen in
Figure~\ref{fig:x_avg}, the ionization fraction averaged across the
slab is sensitive to $u$ but weakly dependent on $T_*$ -- as expected.

\begin{figure}[htp!]  
  \plotone{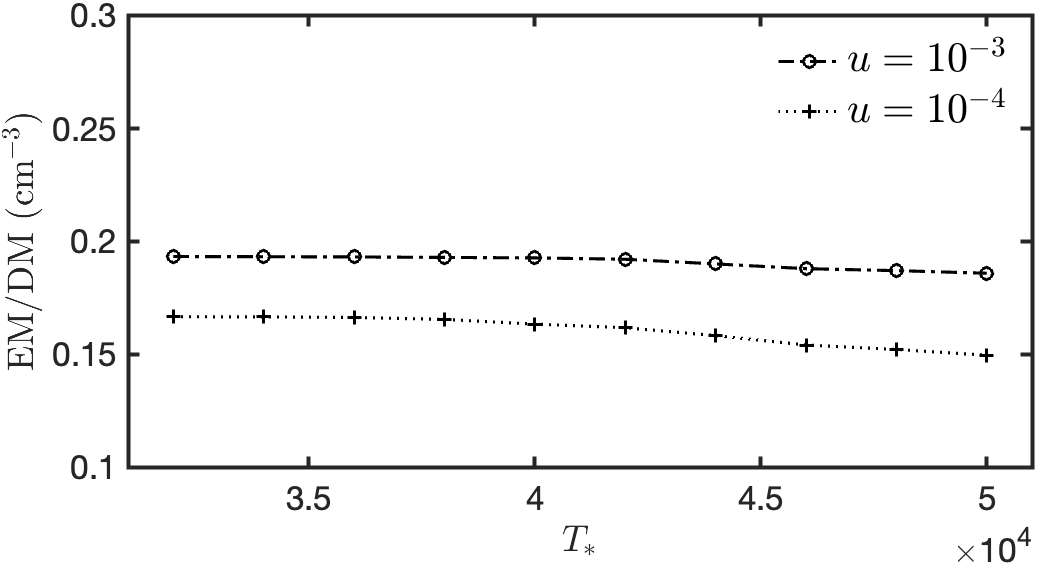}   
   \caption{\small Run of inferred electron density as obtained
   from EM/DM ratio as a function of $T_*$ for two values of $u$.}
 \label{fig:DM_EM} 
\end{figure}

The dispersion measure is given by ${\rm DM}=\int n_edz$ while the
emission measure is given by ${\rm EM}=\int n_en_{\rm H^+}dz$.  A
different manifestation of the partial ionization of the WIM can
be seen in Figure~\ref{fig:DM_EM} where we plot the EM/DM ratio, a
proxy for the mean electron density, as a function of $T_*$.  In
any case, estimating the mass of the WIM based solely on measures
that depend on the electron density (nebular lines, H$\alpha$, DM)
would lead to an underestimate, since the neutral gas
intimately associated with the WIM is not included in the estimate.

\subsubsection{Ionization \&\ Temperature Profile}

\begin{figure*}[htbp]   
 \centering
  \includegraphics[width=3in]{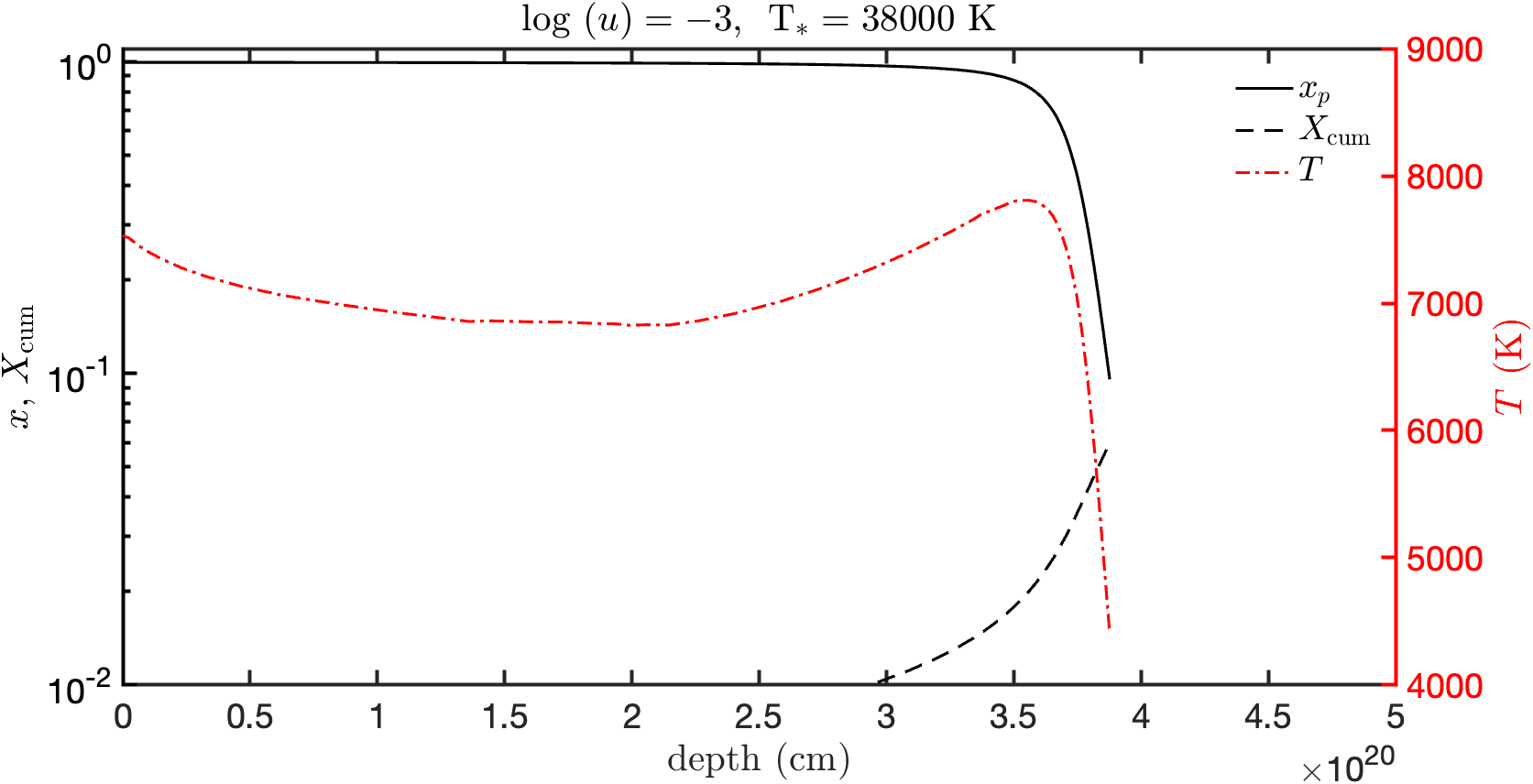}\qquad
   \includegraphics[width=3in]{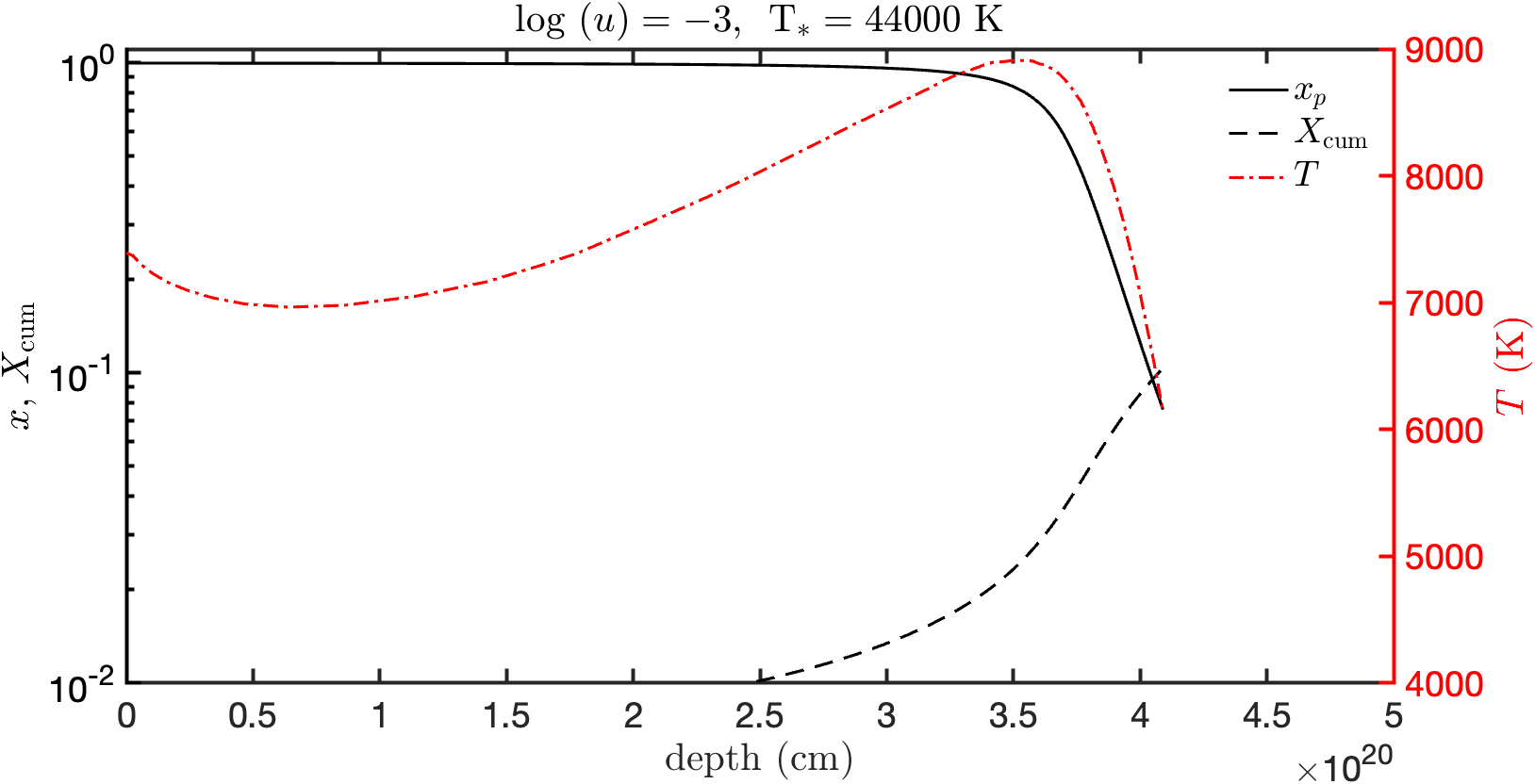} \vskip 0.1in
    \includegraphics[width=3in]{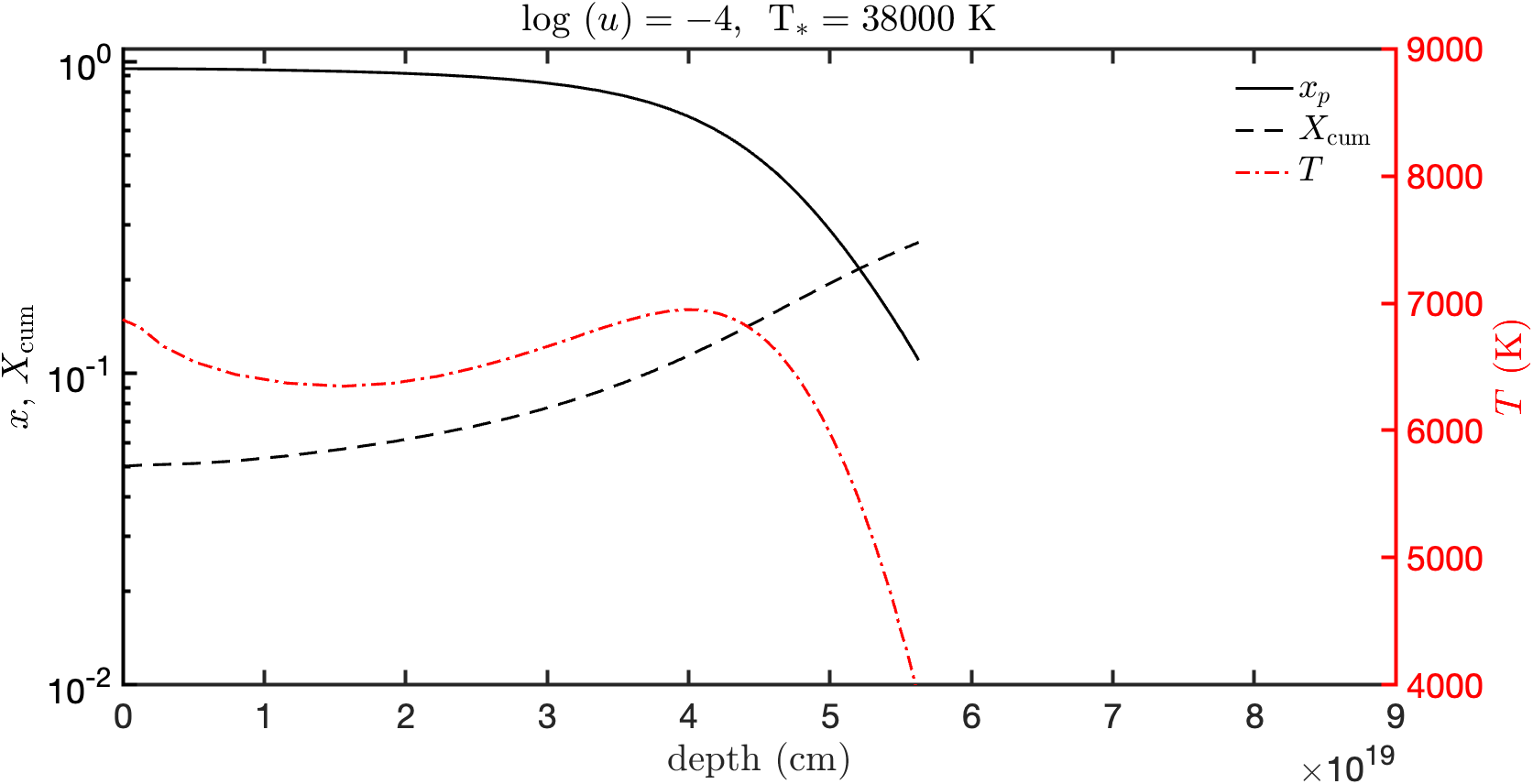}\qquad
    \includegraphics[width=3in]{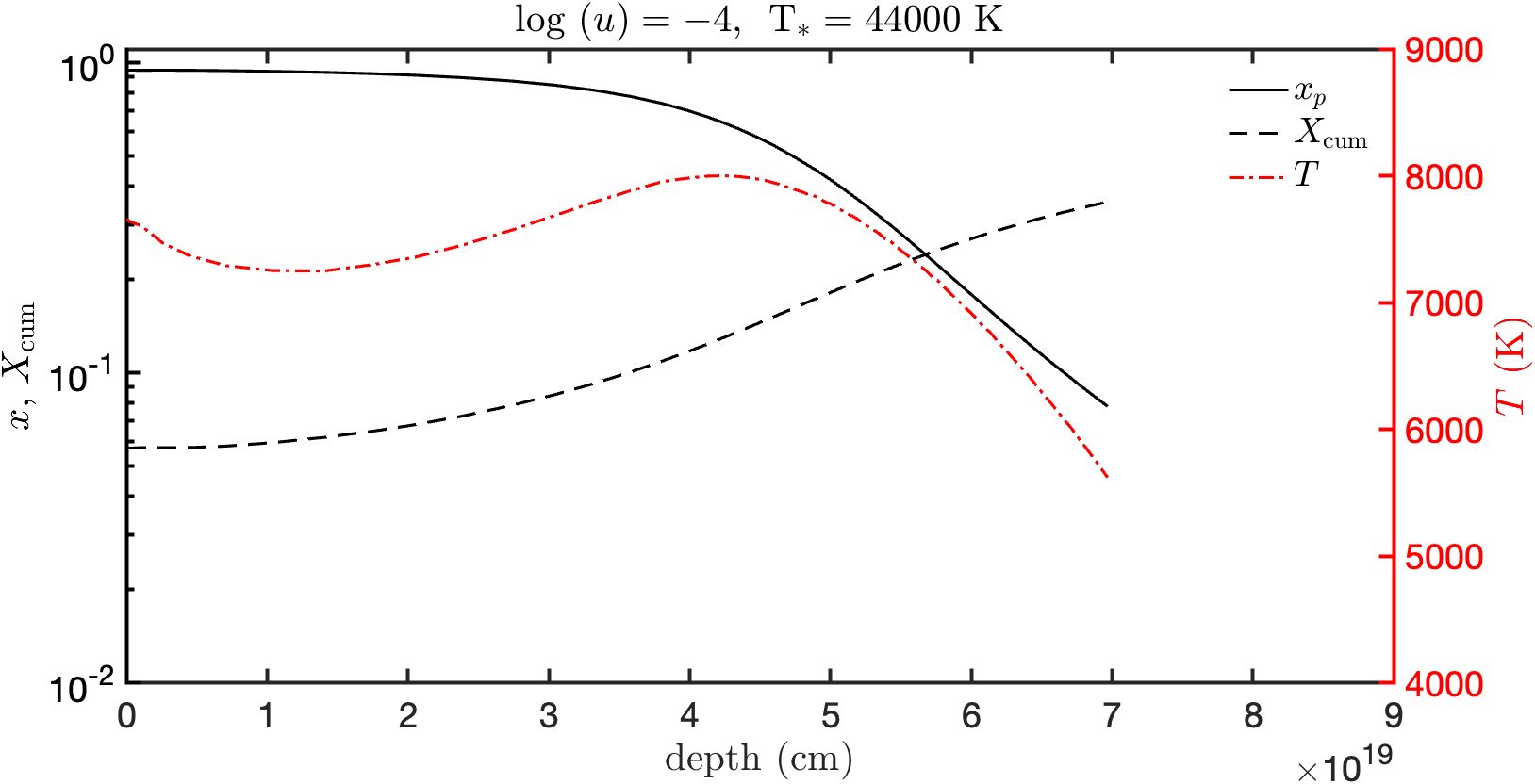}
   \caption{\small Run of gas temperature ($T$; red dot-dash line,
   right vertical axis), ionization of hydrogen ($x_p$;  solid black
   line, left vertical axis) and cumulative neutral fraction of
   hydrogen ($X_{\rm cum}$; black dashed line, left vertical axis)
   as a function of distance into the slab. The calculation was
   terminated when the ionization fraction reached 0.1 at which
   point $z=L$.  The density was fixed to $n_{\rm H}=0.2\,{\rm
   cm^{-3}}$ and the calculations were undertaken for two values,
   each, of $u$ and $T_*$ (noted in the figures).  }
 \label{fig:xT}
\end{figure*}

Here, we review the run of ionization and temperature with depth
for $u=[10^{-3}, 10^{-4}]$ and $T_*=[3.8,4.4]\times 10^4$\,K.  In
Figure~\ref{fig:xT} we plot the run of  electron temperature, the
ionization of hydrogen, $x_p(z)$ and the cumulative neutral fraction,
$X_{\rm cum}(z)$, as a function of depth, $z$.  $x_{\rm cum}(z)=\int_0^z
x dz/\int_0^z dz$ is the cumulative ionization fraction while $X_{\rm
cum}(z)\equiv 1-x_{\rm cum}(z)$.

We expect DM to increase linearly with $u$ (larger $L$). We expect
EM to increase linearly with $u$ (due to larger $I_0$; see
\S\ref{sec:1-D}).  In fact, these expectations are borne out by the
CLOUDY model.  For $u=10^{-4}$, DM increases with $T_*$ from
2.6\,cm$^{-3}$\,pc to 3.2\,cm$^{-3}$\,pc while EM increases from
0.43\,${\rm cm^{-6}\,pc}$ to 0.48\,${\rm cm^{-6}\,pc}$.  The
corresponding values for $u=10^{-3}$ are [23.4, 25.4]\,cm$^{-3}$
[4.52,4.73]\,cm$^{-6}$\,pc, respectively.

\subsubsection{Line Emission}

\begin{figure}[htbp] 
 \plotone{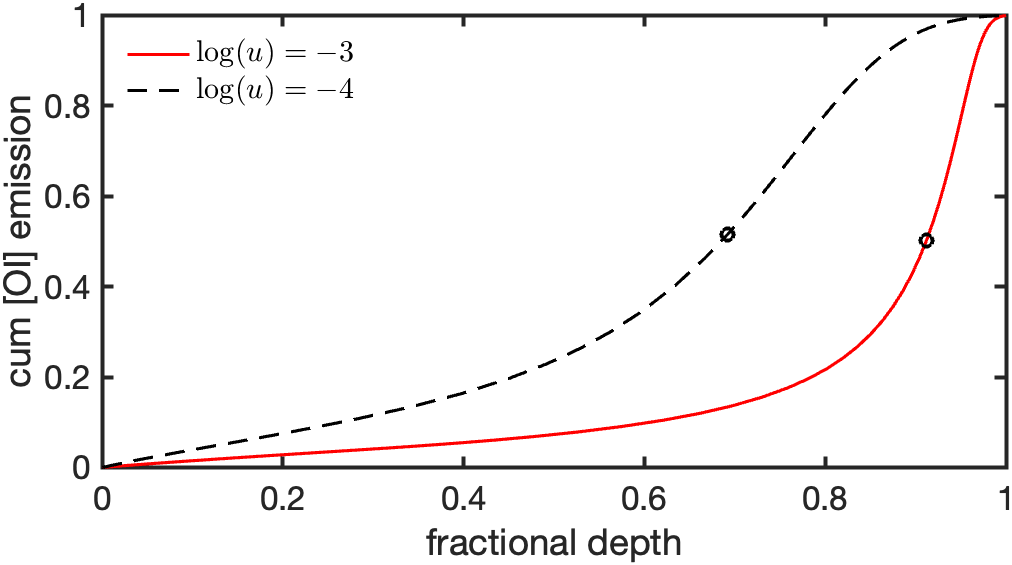} 
  \caption{\small The run of normalized cumulative [OI] emission against
   fraction depth, $z/L$, for two values of $u$ and $T_*=38,000$\,K.
   Here, $L$ is the depth at which $x=0.1$ -- the edge 
   of the WIM (see caption to Figure~\ref{fig:xT}). Let $C(z)$
   be the [OI] emission per unit volume. The $y$-axis is given by
   $\int_0^zC(z)/\int_0^L C(z)dz$.  The two circles mark an ordinate
   value of 0.5. }
 \label{fig:OI_depth}
\end{figure}

Assuming {\it (i)} O$^+$\!/O$^0$=(8/9)H$^+$\!/H$^0$ and {\it (ii)}
electron collisions approximated by hard spheres, we compute emission of [OI]\,6300 per unit volume.  We plot, in Figure~\ref{fig:OI_depth},
the run of the modeled emission of [OI]\,6300 as a function of the
fractional depth, $z/L$.  We can see that for $u=10^{-4}$, the [OI]
emission is from most of the column. In contrast, for $u=10^{-3}$,
the emission arises primarily in the transition region.

\begin{figure}[htbp]   
 \plotone{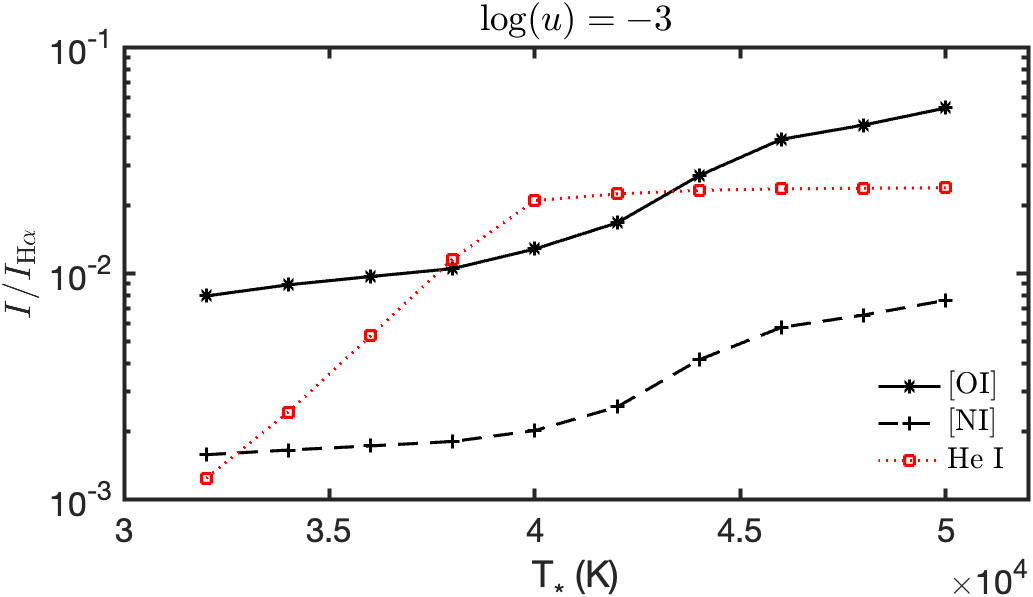}\vskip 0.2in
  \plotone{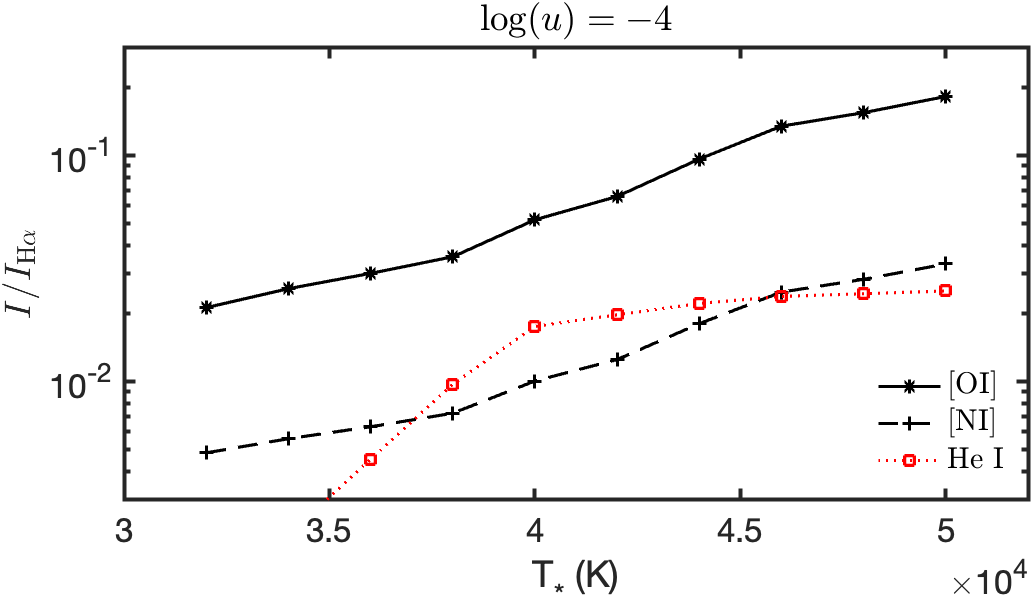}
   \caption{\small Run of intensity of
    [OI] 6300, [NI] 5200 and He~I 5875  relative to the H$\alpha$
    intensity  with $T_*$. The parameters for CLOUDY are: $n_{\rm
   H}=0.2\,{\rm cm^{-3}}$ and $u=[10^{-3}, 10^{-4}]$.  }
    \label{fig:OI_NI}
\end{figure}

In Figure~\ref{fig:OI_NI} we display the CLOUDY-computed [OI] and
[NI] intensities\footnote{CLOUDY uses line-integrated intensity in
energy units, ${\rm erg\,cm^{-2}\,s^{-1}\,ster^{-1}}$ whereas
observers prefer photon intensity carrying the unit of ${\rm
phot\,cm^{-2}\,s^{-1}\,ster^{-1}}$.  We denote the energy intensity
by $I$ and use $\mathcal{I}$ for photon intensity.} normalized to
the value of H$\alpha$ as a function of $T_*$. In Figure~\ref{fig:OI_NI}
we see that $I_{\rm OI}/I_{\rm H\alpha}$ and $I_{\rm NI}/I_{\rm
H\alpha}$ increase with decreasing $u$, as expected.  The electron
temperature increases with $T_*$.  The emission rapidly increases
with $T$ (thanks to the Boltzmann factor).  This reasoning explains
the rapid strengthening of line emission with $T_*$. Note that the
$I/I_{\rm H\alpha}$ shown in Figure~\ref{fig:OI_NI} were obtained
assuming that the WNM slab was thick enough to accommodate the
transition from WIM to WNM.  The line ratios will be lower if we
demanded $X_{\rm edge}$ to be small, say 0.5 or so.

\cite{shr+00} present CLOUDY calculations for $T_*=38,000\,$K and
$u=10^{-4}$.  In Table~4 of their paper they report $I_{\rm OI}/I_{\rm
H\alpha}=0.051$ while \cite{dm94} find 0.055.  For the same $T_*$
and $u$ we find $I_{\rm OI}/I_{\rm H\alpha}=0.036$.  The strength
of the [OI] line depends linearly on the abundance of oxygen but,
as noted above, will increase rapidly with the electron temperature.
The latter is determined by a balance between heating and cooling,
which is primarily due to nebular lines of metals. An increased
abundance of metals leads to a cooler WIM.  Compared to \cite{dm94}
and \cite{shr+00}, the metal abundances are systematically higher
in our model. As a result, for the same values of $T_*$ and $u$,
the mean electron temperature in our CLOUDY model is 6,060\,K
(Figure~\ref{fig:Tstar_Tgas}), significantly lower than 8,040\,K
\citep{shr+00} and 8,200 \citep{dm94}.  So, it is not unreasonable
that our value for $I_{\rm OI}/I_{\rm H\alpha}$ is smaller than
that of \cite{shr+00}.  We will return to the important issue of
metal abundance of the WIM in the concluding section
(\S\ref{sec:SummaryConclusions}).

\section{Optical and NIR lines of [NI]}
  \label{sec:N_Lines}
 
\begin{figure}[hbtp]
 \centering
 \includegraphics[width=2.3in]{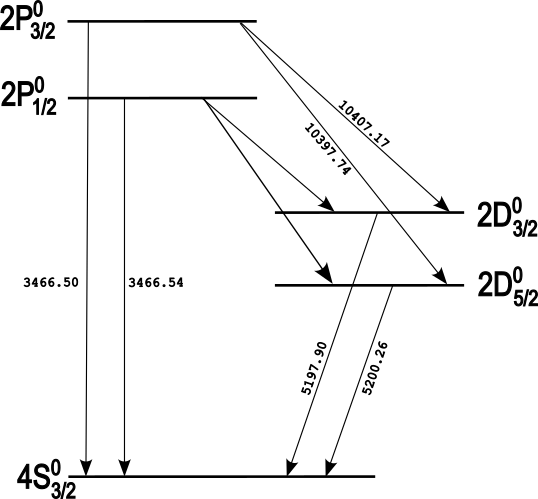}
  \caption{\small Grotrian diagram of N~I  for ground and first two
  excited terms, all with the same electronic configuration
  ($1s^22s^22p^3$). To avoid clutter, the wavelengths for a pair
  of lines resulting from $^2{\rm P}^{\rm o}_{\sfrac{1}{2}} \rightarrow
  {}^2{\rm D}^{\rm o}$ are not shown. See Table~\ref{tab:NI_Atomic}
  for additional details.  Incidentally the transition, $^2{\rm
  D}^{\rm o}_{5/2}$--$^2{\rm P}^{\rm o}_{1/2}$, with  $\Delta J=2$
  is doubly forbidden.  Nonetheless, as can be gathered from
  Table~\ref{tab:NI_Atomic}, the A coefficient of this line is
  comparable to other lines of the quartet.  } 
	\label{fig:NI_Grotrian}
\end{figure}

The Grotrian diagram for N~I is presented in Figure~\ref{fig:NI_Grotrian}.
By tradition, line transitions resulting from the decay of the
levels in the first exited term to the ground term are referred to
as ``nebular", those that result from the decay of the second excited
term to the first excited term are ``auroral" and finally those
that decay from the second excited term to the ground term are
``trans-auroral" \citep{bmp33}.  Thus, with reference to
Table~\ref{tab:NI_Atomic}, the V-band doublet
[NI]\,$\lambda\lambda$5200.3,\,5197.9 is nebular, the quartet
centered around  1\,$\mu$m is auroral and the  U-band doublet
[NI]\,$\lambda\lambda$3466.50,\,3466.54 is trans-auroral.

\subsection{The nebular lines of N~I}
 \label{sec:NebularLines_NI}
 
We assume excitation through collisions, specifically electrons
(number density, $n_e$) and ignore stimulated emission.  The
excitation rate per unit volume is $n_en_l q_{lu}$ where $n_l$ is
the number density of nitrogen atoms in the lower (``$l$") state
and $q_{lu}$ is the collision excitation coefficient from $l$ to
the upper state, $u$.  The collisional excitation coefficient is
given by
 \begin{equation}
   q_{lu} = \frac{8.629\times 10^{-8}}{T_4^{1/2}}
    \frac{\Omega_{ul}}{g_l} {\rm e}^{-E_{ul}/k_BT}\ {\rm cm^3\,s^{-1}}
    \ .  \nonumber
   \end{equation}
Here, $\Omega_{ul}$ is the collision strength, $E_{ul}$ is the line
energy, $T=10^4T_4\,$ K is the electron temperature, and $g_l$
($g_u$) is the degeneracy of the lower level (upper level).  The
corresponding collisional de-excitation coefficient is $q_{ul}$ and
is related to $q_{lu}$ through a detailed balance (see \S2.3 of
\citealt{D11}).  The collision strengths and A coefficients ($A_{ul}$)
of the relevant lines of N$^0$ can be found in Table~\ref{tab:NI_Atomic}.
Additional collision strengths can be found in Appendix~F of
\citet{D11}.  In steady state, the ratio of the density of atoms
in the upper level to those in the lower level is given by
$(n_u/g_u)/(n_l/g_l) ={\rm exp}(-E_{ul}/k_BT)/(1+n_{\rm cr}/n_e)$
where the ``critical" density is $n_{\rm cr}= A_{ul}/q_{ul}$  (see
\S 17.2 in \citealt{D11}).

\begin{deluxetable}{llr}[hbt]
 \tablecaption{Forbidden lines of N~I}
\label{tab:NI_Atomic}
\tablewidth{0pt}
\tablehead{
 \colhead{$l$-$u$} & 
 \colhead{$\lambda$ (\AA)}  &
 \colhead{$A_{ul}\,({\rm s}^{-1})$} }
\startdata
$^4{\rm S}_{\sfrac{3}{2}}^{\rm o}$-$^2{\rm D}_{\sfrac{5}{2}}^{\rm o}$ & 5200.26 &
 $7.56\times 10^{-6}$ \\
$^4{\rm S}_{\sfrac{3}{2}}^{\rm o}$-$^2{\rm D}_{\sfrac{3}{2}}^{\rm o}$ & 5197.90&  
$2.03\times 10^{-5}$ \\
\hline
$^4{\rm S}_{\sfrac{3}{2}}^{\rm o}$-$^2{\rm P}_{\sfrac{3}{2}}^{\rm o}$ & 3466.50&
 $6.5\times 10^{-3}$\\
$^4{\rm S}_{\sfrac{3}{2}}^{\rm o}$-$^2{\rm P}_{\sfrac{1}{2}}^{\rm o}$ &3466.54 &  
$2.6\times 10^{-3}$ \\
\hline
$^2{\rm D}_{\sfrac{3}{2}}^{\rm o}$-$^2{\rm P}_{\sfrac{3}{2}}^{\rm o}$& 10407.17&
$2.7\times 10^{-2}$ \\
$^2{\rm D}_{\sfrac{3}{2}}^{\rm o}$-$^2{\rm P}_{\sfrac{1}{2}}^{\rm o}$& 10407.59&
$5.3\times 10^{-2}$ \\
$^2{\rm D}_{\sfrac{5}{2}}^{\rm o}$-$^2{\rm P}_{\sfrac{3}{2}}^{\rm o}$& 10397.74&
$6.1\times 10^{-2}$\\
$^2{\rm D}_{\sfrac{5}{2}}^{\rm o}$-$^2{\rm P}_{\sfrac{1}{2}}^{\rm o}$& 10398.15&
$3.4\times 10^{-2}$\\
\enddata
\tablecomments{\small $l$ and $u$ stand for lower and upper states;
 $\lambda$ is the air wavelength; and $A_{ul}$ is he A coefficient
 for transition $u\rightarrow l$.  The data are from NIST. For
 $\lambda$5198 and $\lambda$5200 we have summed the A-coefficients
 of the M1 and E2 transitions.  The accuracy of the A-coefficients
is graded as ``B" in the NIST database.  The collision strengths
 for the two nebular lines are $\Omega_{ul}$, are $0.337
 T_4^{0.732-0.129{\rm ln}(T_4)}$ and $0.224 T_4^{0.726-0.125{\rm
 ln}(T_4)}$, respectively (from Appendix~F of \citealt{D11}).   }.
\end{deluxetable}

At $T=8,000\,$K, we find $n_{\rm cr}=1.7\times 10^3\,{\rm cm^{-3}}$
and $4.4\times 10^3\,{\rm cm^{-3}}$ for [NI]\,5200 and [NI]\,5198,
respectively.  In the WIM, $n_e\ll n_{\rm cr}$ and so the level
population becomes sub-Boltzmann with the result that the photon
intensity is $\propto \Omega_{ul}/g_l$.  However, as we will see
later, strong [NI] and [OI] arise in the ionosphere where $n_e\gg
n_{\rm cr}$.  In this case,  the level population assumes the
Boltzmann distribution and, as a result, the photon intensity is
$\propto A_{ul}g_u$.

Parenthetically, we note\footnote{p.\ 52 of \cite{of06}} the
following: for transitions involving a term with a single level
($L=0$ or $S=0$) and a term with multiple levels, the collision
strengths for the multiple levels are $\propto (2J^\prime+1)$ where
the prime refers to levels of the term with multiple levels. Thus,
in the limit of $n_e\ll n_{\rm cr}$ and assuming that only collisions
excite the atoms from the ground state (${\rm ^4S_{3/2}^{\rm o}}$)
to the first excited term (${\rm ^2D^{\rm o}}$), the photon intensity
ratio is $\beta_W\equiv I_{5198}/I_{5200}=2/3$.

We assume that nitrogen is not depleted onto grains.  We set the
relative abundance of nitrogen to that of hydrogen, by number, to
74\,ppm (parts per million; Chap. 1 of \citealt{D11}).  Using the
collision strength given in Table~\ref{tab:NI_Atomic} we find that
the photon intensity is given by
 \begin{equation}
  \mathcal{I}_{5200} = 1.66 \bigg(\frac{x_{\rm N^0}}{x_{\rm H^+}}\bigg)
  T_4^\gamma {\rm EM}\ {\rm e}^{-2.766/T_4} \ R
   \label{eq:I5200}
 \end{equation}
where $x_{\rm N^0}=n({\rm N^0})/n_{\rm N}$ is the neutral fraction
of nitrogen; $x_{\rm H^+}=n_{\rm H^+}/n_{\rm H}$ is the ionized
fraction of hydrogen; EM is the emission measure, $\int n_en_{\rm
H^+}dl$, which carries the unit of ${\rm cm^{-6}\,pc}$; $R$ represents
Rayleigh, a unit of photon intensity numerically equal to
$10^6/(4\pi)\,{\rm photon\,cm^{-2}\,s^{-1}\,ster^{-1}}$ and
$\gamma=0.223-0.129{\rm ln}(T_4)$. For $T=8,000\,$K, we find
$\mathcal{I}_{5200}=0.082(x_{\rm N^0}/x_{\rm H^+})\,{\rm EM}\,R$.

There is only one published Fabry-P\'erot observation of the diffuse
ISM in the [NI]\,$\lambda$5200 line \citep{rrs77}.  The authors
place a limit, $\mathcal{I}_{5200}/\mathcal{I}_{\rm H\alpha}<7\times
10^{-3}$.

\subsection{Auroral and trans-auroral lines of N~I}
	\label{sec:Auroral}
	
Nitrogen atoms excited to the second excited term, $^2{\rm P}^{\rm
o}$, can decay by producing the U-band doublet or by producing the
1-$\mu$m quartet.   Using the A coefficients listed in
Table~\ref{tab:NI_Atomic} we find the branching fraction to be 6.8\%
for [NI]\,3466.50 and 2.9\% for [NI]\,3466.54.  The integrated intensity
of the 1-$\mu$m quartet is given by the sum of the excitations from
the ground state to the $^2{\rm P}^{\rm o}$ term (both levels) times
the branching probability ($\approx 0.9$),
 \begin{equation*}
    \mathcal{I}_{\rm 1\mu m} \approx 9.9 \Big(\frac{x_{\rm N^0}}{x_{\rm
    H^+}}\Big) T_4^{0.26-0.14{\rm ln}(T_4)} {\rm e}^{-4.151/T_4}\,
    {\rm EM}\,R\ .
   \label{eq:I_1micron}
 \end{equation*}
As before, we set $T=8,000\,$K and find $\mathcal{I}_{\rm 1\mu{m}}\approx
0.052 (x_{\rm N^0}/x_{\rm H^+}){\rm EM}\,R$.

The region around 1\,$\mu$m is infested by bright OH Meinel bands
(see \citealt{rlc+00} for high spectral resolution spectra of the
night sky).  So we have no choice but to turn to a space-based
facility which, assuming a mission at 1\,AU, is limited by zodiacal
light.  The background, even at the ecliptic poles, is 23.3
mag\,arcsec$^{-2}$ in the V-band \citep{lbh+98}.  This corresponds
to $1.74\,\mu{\rm Jy\,arcsec^{-2}}$ or $0.14\,R\,{\textrm \AA}^{-1}$.
A Fabry-P\'erot spectrometer or a high-resolution imaging spectrograph,
$R_\lambda\gtrsim 10^4$ operating in space, can probe
auroral lines of N$^0$.

\subsection{Doublet Ratio: WIM}
 \label{sec:DoubletRatio_WIM}
 
The optical doublet [NI]\,$\lambda\lambda$5198,\,5200 results from
the decay of the first excited term (${\rm ^2D^{\rm o}}$) to the
ground term (see Figure~\ref{fig:NI_Grotrian}). The first excited
term can be reached by collisional excitation from the ground state
(discussed in \S\ref{sec:NebularLines_NI}) or by decay from the
${\rm ^2P^{\rm o}}$ term (discussed in \S\ref{sec:Auroral}).  In
the low-density limit, $n_e\ll n_{\rm cr}$, every collision results
in radiative decay.  We have calculated the intensity ratio $
\mathcal{I}_{5198}/\mathcal{I}_{5200}$ that includes contributions
from excitation to the first and second excited terms.  In the
temperature range of interest, this ratio is, within 1\%, the same
as that strictly computed for the nebular decay, $\beta_W= 2/3$
(see \S\ref{sec:NebularLines_NI}).  The reason for this constancy
is that decays of ${\rm ^2P^{\rm o}}$  feed the two levels  ${\rm
^2D_{3/2}^{\rm o}}$ and ${\rm ^2D_{5/2}^{\rm o}}$ with a ratio of
0.6933  which is close to the 2/3 derived in  \S\ref{sec:NebularLines_NI}
for only collisional excitations to ${\rm ^2D^{\rm o}}$ term.  So,
we set $\beta_W=2/3$.

\section{[NI] Airglow}
 \label{sec:Airglow_NI}

\begin{figure*}[hbtp]     
 \centering
  \includegraphics[width=0.8\textwidth]{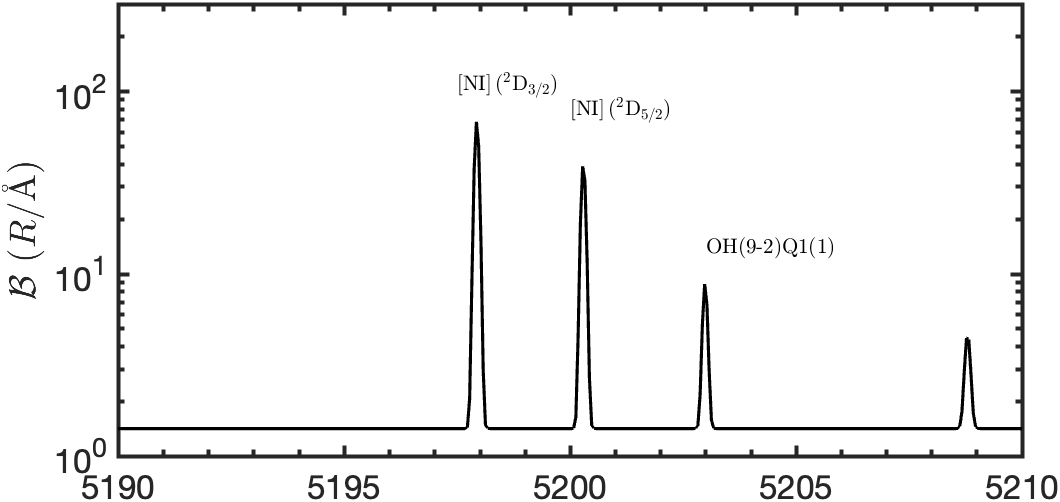}
   \caption{\small Graphical display of the flux-calibrated sky
   intensity (Rayleigh per \AA) centered around the [NI] nebular
   doublet. The strongest airglow line next to the doublet is marked
   as such.  The spectrum was constructed from the sky line catalog
   of \citet{H03} and \citet{css+06}.  Notice the log scaling for
   the ordinate.   Consistent with the mean spectrum of \citet{H03},
   the displayed continuum is about $10^{-17}\,{\rm erg\,
   cm^{-2}\,s^{-1}\,arcsec^{-2}}\,$\AA$^{-1}$ or about
   1.5\,$R$\,\AA$^{-1}$. }
 \label{fig:ESO_Sky_Spectrum_Nitrogen}
\end{figure*}

Most astronomers regard emission by the Earth's atmosphere (airglow,
aurora) to be a nuisance. Those who are studying other galaxies
carefully choose the redshift range to avoid bright atmospheric
emission lines. However, astronomers interested in the study of the
Galactic ISM have no choice\footnote{In fact, this paper was only
possible due to partnership between astronomers and atmospheric
scientists.} but to accept airglow! One of the authors (SRK)
recommends the following reading list for astronomers who wish to
have a basic understanding of [OI] and [NI] emission from the
atmosphere.  \cite{cnh16} provides a gentle overview.  \cite{B78}
provides a very readable and scholarly history of the recognition
and development of our understanding of the [OI] and [NI] lines.
The retrospective paper on the {\it Visible Airglow Experiment}
(VAE) by \cite{has+88} is an excellent starting point for those
wishing to understand the underlying physics and chemistry of [OI]
and [NI] airglow lines.  See \cite{nkb+12} for a summary of atmospheric
emission specific to the Paranal site.

\citet{H03} presents high spectral resolution ($R_\lambda \approx
45,000$ in the blue and 43,000 in the red) sky spectra taken with
Ultraviolet \&\ Visual Echelle Spectrograph (UVES; \citealt{ddk+00})
at ESO's VLT facility in Chile, while \cite{css+06}
identified the lines.  In Figure~\ref{fig:ESO_Sky_Spectrum_Nitrogen}
we show the sky spectrum in the vicinity of the [NI] doublet.

\begin{figure}[htbp] 
 \plotone{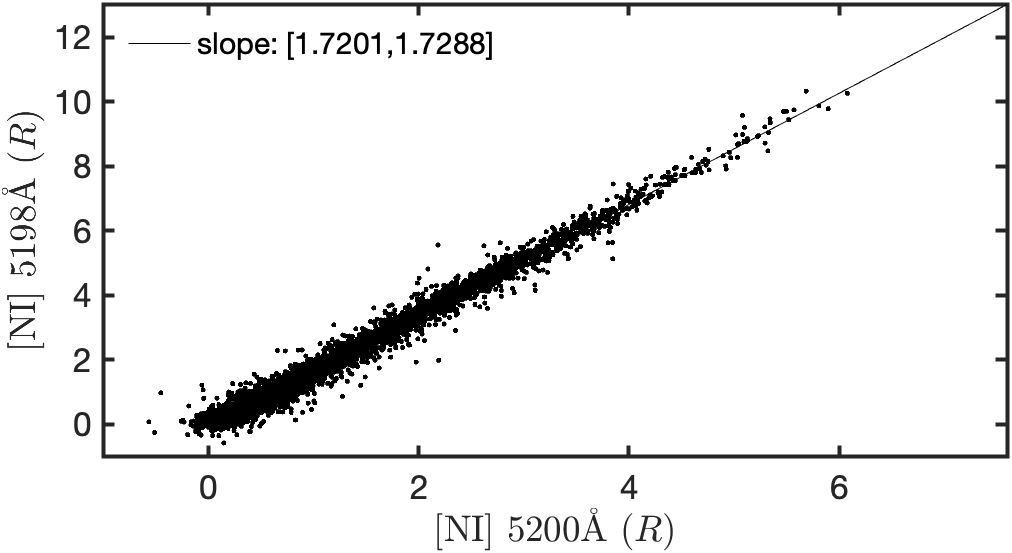}
  \caption{\small The scatter plot of intensities of [NI]\,5198
  versus [NI]\,5200.  The two intensities are highly correlated.
  The 95\% confidence interval for the slope of the best fit linear
  model is displayed in the figure.}
 \label{fig:NI_ratio}
\end{figure}

Given the 2.4-\AA\ separation for the V-band doublet, we chose to
focus on archival data obtained from X-shooter, a versatile
intermediate resolution near-UV to near-IR spectrograph mounted on
one of the VLT 8.2-m telescopes \citep{vdd+11}.  Since the start
of regular X-shooter observations in 2009, many slit spectra
with a significant sky contribution have been collected. This large
data set has already been used for airglow research
\citep{noll22,noll23,noll24}.  The details of the procedure are
described in \S\ref{sec:X-shooter}.

The main driver of [NI] emission in the Earth's atmosphere are
$\rm{O}^{+}$ ions which are produced by solar UV photons at daytime
in the ionospheric F layer at an altitude of several hundred
kilometers.  According to the International Reference Ionosphere
model (IRI; see \citealt{bilitza22}), the peak of ion and
electron density, at night, is mostly between 250\,km and
350\,km in Paranal.  The $\rm{O}^{+}$ ions can react with nitrogen
molecules.  The charge transfer reaction results in $\rm{NO}^{+}$ and
N atoms (e.g., \citealt{kelley09}).  Finally, dissociative electron
recombination leaves N$^0$ atoms in the $^2{\rm D}^{\rm o}$ term.
As the production of excited N atoms involves $\rm{N}_2$ molecules,
the effective height of the [NI] emission is lower than the peak
of the F layer.  \citet{ssc+05} report a rough altitude of 200\,km.

\subsection{Doublet Ratio: Airglow}
 \label{sec:Airglow_DoubletRatio}

From Figure~\ref{fig:NI_ratio} we see that the airglow doublet lines
of [NI] are strongly correlated. A linear model provides an excellent
fit, $\mathcal{I}_{5198}=\beta_A \mathcal{I}_{5200}$ where
$\beta_A=[1.720,1.729]$ (95\% confidence).  This ratio is inverted
from that expected in the WIM (see \S\ref{sec:DoubletRatio_WIM}).
A representative electron density at a height of 200\,km is
$10^{10}\,{\rm cm^{-3}}$ (e.g. \citealt{emmert15}).  For a
thermospheric temperature of about $10^3\,$K (cf.\ \citealt{jacchia71})
the critical density for the [NI] doublet is about $10^3\,{\rm
cm^{-3}}$ (see \S\ref{sec:NebularLines_NI}).  Thus, in the upper
atmosphere where nitrogen atoms reside, the level population follows
the Boltzmann distribution, and so the intensity is $\propto A_{ul}g_u$
(see \S\ref{sec:NebularLines_NI}).  Using $A_{ul}$ given in
Table~\ref{tab:NI_Atomic}, in this limit, we expect
$\mathcal{I}_{5198}/\mathcal{I}_{5200}=1.793$.

Apparently, airglow line intensity ratios can be measured with a
higher precision than our ability to compute A coefficients. This
fact has motivated aeronomers to undertake precision measurements
of several doublets.  \citet{ssc+05}, using data from the Keck
Observatory (HIRES and ESI), report a value of $1.759\pm 0.014$.
Our 95\% confidence interval for the slope (Figure~\ref{fig:NI_ratio})
is in marginal agreement with the Keck value. Neither of these
agrees with that calculated using the A coefficients reported by
NIST.  We set $\beta_A\approx 1.74$, the mean of the Keck and ESO
values.

\begin{figure}[htbp]   
 \plotone{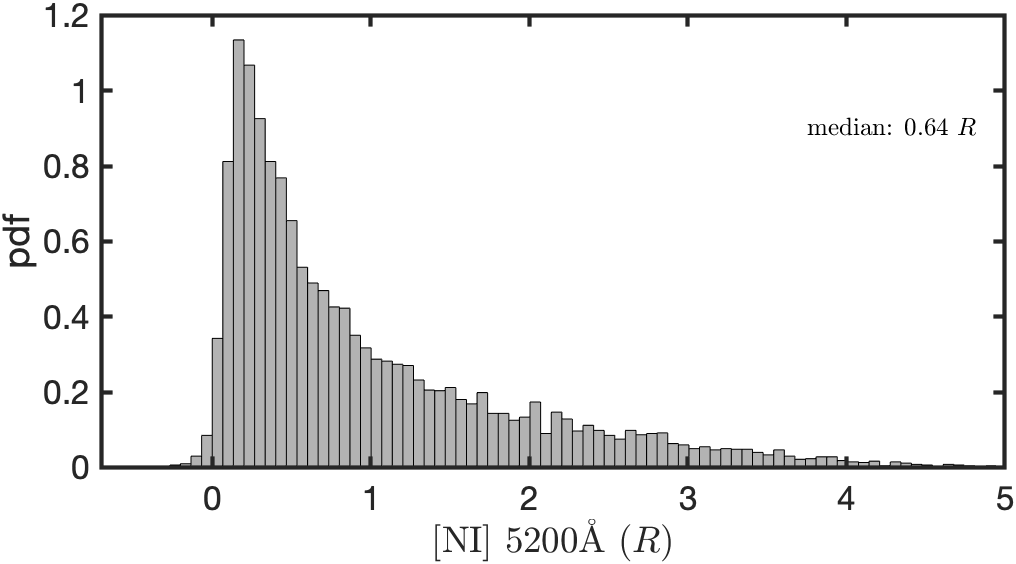}
  \caption{\small The histogram of the intensity of the [NI]\,5200
  airglow line (extrapolated to the zenith).}
 \label{fig:Nitrogen_Histogram}
\end{figure}

A histogram of the intensity of the [NI]\,5200 line can be found in
Figure~\ref{fig:Nitrogen_Histogram}.  The median value is 0.64\,$R$
which is two orders of magnitude smaller than the strength of the
[OI]\,6300 airglow line (\S\ref{sec:OI_Airglow}).  The [NII]\,5200
airglow can be compared with the typical continuum\footnote{The
continuum is a mixture of airglow emission (see \citealt{noll24})
and other sky radiation components (zodiacal light, scattered
moonlight, scattered starlight; see \citealt{nkb+12}) and can be
quite variable.} of $1R\,$\AA$^{-1}$ and  a geocoronal H$\alpha$
intensity of 2--4\,$R$ for typical night shadow heights of the order
of the Earth radius \citep{nmr+08}.

\subsection{Optimizing for observations of [NI]}
 \label{sec:Optimizing}
 
\begin{figure}[htbp]    
 \plotone{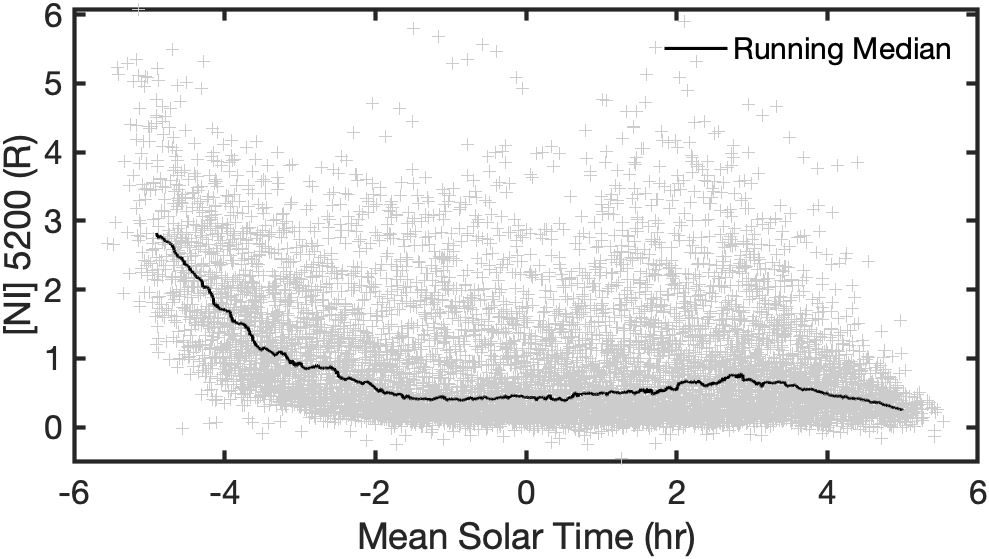}
  \caption{\small Moving (boxcar length: 300 points) median of
  [NI]\,5200 emission as a function of the mean solar time (MST).
  The background gray cloud is the collection of data points which
  are plotted as gray points.  The mean solar time is approximately
  local solar time. The longitude of the Cerro Paranal site is
  70.4$^\circ$ West. MST is defined to be UT $-4.7$\,hr.  }
 \label{fig:Airglow_MST}
\end{figure}

In Figure~\ref{fig:Airglow_MST} we see that the median intensity
of the [NI] airglow decreases as the night progresses (with the
exception of a shallow local maximum near 3\,am). Thus, observations
should ideally be made after 9 to 10\,pm (local time). Even so,
there are still large variations. As the X-shooter data show, some
of the variations can be attributed to strong seasonal intensity
variations with maxima around the equinoxes; and also a strong
positive dependence on solar activity (e.g., \citealt{nkb+12}), as
traced by the solar radio flux \citep{tapping13}. In any case, the
electron and ion densities in the upper atmosphere are key to the
measured airglow intensities. As ionization occurs mainly at the
beginning of the day, a nocturnal decrease is expected.  However,
the actual behavior is also influenced by large-scale transport
processes, instabilities, and wave propagation \citep[e.g.,][]{kelley09}.
The airglow intensity also increases with increasing zenith angle
as a result of the thicker projected width of the emission layer.
For a zenith angle of 60$^{\circ}$ and an emission altitude of
200\,km, the intensity is 1.84 times higher than at the zenith
\citep{vanrhijn21}. As discussed in \S\ref{sec:X-shooter}, the data
plotted in this and other such figures have been corrected for this
effect and are therefore representative of the zenith, which would
be the optimum direction for extraterrestrial [NI] observations.

For the WIM, the expected signal strength of [NI]\,5200 is well below
$1\,R$ which means that any realistic measurement must take into
account airglow and continuum emission. The ionospheric
airglow lines arise in regions where nighttime temperatures
often range between 600\,K and 1,100\,K
\citep[e.g.,][]{emmert15,jacchia71}.  The thermal full width at
half-maximum (FWHM) of the [NI] line is narrow, $1.8\,{\rm
km\,s^{-1}}$ for 1,000\,K.  On the other hand, the thermal FWHM of
the WIM [NI] line, assuming a temperature of 8,000\,K, is
5.1\,km\,s$^{-1}$. The airglow line has a topocentric velocity of
0\,km\,s$^{-1}$ while the WIM lines can be offset by $\pm 30\,{\rm
km\,s^{-1}}$ (motion of Earth relative to the Solar System barycenter),
$\pm 15\,{\rm km\,s^{-1}}$ (Local Standard of Rest relative to the
barycenter) and the Galactic rotation curve.

The simplest case is when there is wide velocity separation between
the airglow lines and the WIM lines. This approach then requires
spectrometers with very high spectral resolution $R_\lambda\gtrsim
3\times 10^4$.  Lacking such high spectral resolution, we can use
the doublet method described below to disentangle the WIM contribution
from the airglow.

\subsection{The Doublet Method}
 \label{sec:NI_DoubletMethod}
 
\begin{figure}[htbp]    
 \plotone{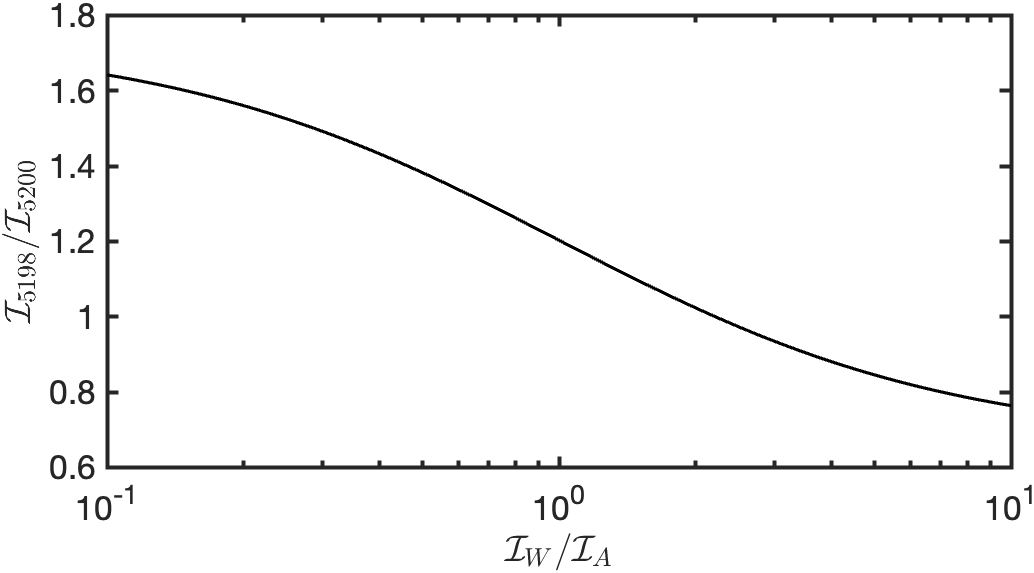}
  \caption{\small The run of the observed doublet ratio,
  $\mathcal{I}_{5200}/\mathcal{I}_{5198}$, as a function of the
  ratio of the [NI]\,5200 photon intensity from the WIM ($\mathcal{I}_W$)
  to that of the airglow ($\mathcal{I}_A$); see
  Equation~\ref{eq:mathcal_R}. In the limit where airglow dominates
  the doublet ratio is closer to $\beta_A\approx 1.74$ while in the
  opposite limit it is closer to $\beta_W=2/3$.} \label{fig:NI_WIM_Airglow}
\end{figure}

Let $\mathcal{I}_A$ ($\mathcal{I}_W$) be the photon intensity of
[NI]\,5200 line arising from the airglow (WIM).  The intensity of
[NI]\,5198 airglow line is then  $\beta_A\mathcal{I}_A$
(\S\ref{sec:DoubletRatio_WIM}) while the emission from the WIM is
$\beta_W\mathcal{I}_W$ (\S\ref{sec:Airglow_DoubletRatio}).  Thus,
the measured doublet ratio is
 \begin{equation}
  \frac{\mathcal{I}_{5198}}{\mathcal{I}_{5200}} =
	\frac{\beta_A+\beta_W(\mathcal{I}_W/\mathcal{I}_A)}{1+
	(\mathcal{I}_W/\mathcal{I}_A)}
		\label{eq:mathcal_R}
 \end{equation}

The run of the observed $\mathcal{I}_{5198}/\mathcal{I}_{5200}$
versus the ratio of [NI]\,5200 emission from the WIM to that of
the airglow is plotted in Figure~\ref{fig:NI_WIM_Airglow}. The 
measured doublet ratio can yield information on the WIM contribution,
albeit with reduction in signal-to-noise ratio that varies with the
intensity of the airglow.

\section{Nebular and Airglow lines of O~I}
 \label{sec:OI}

\begin{figure*}[htbp]   
 \centering
  \includegraphics[width=5in]{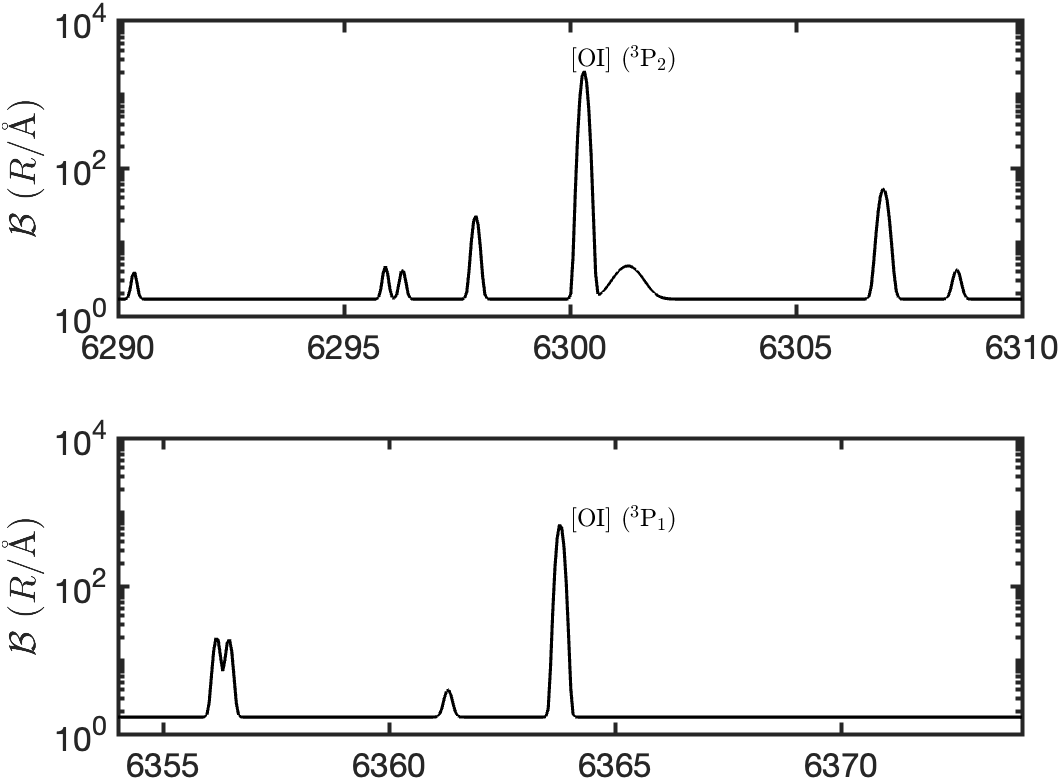}
   \caption{\small The spectrum of the sky in the vicinity of the
   [OI] doublet constructed in the same manner as described in the
   caption of Figure~\ref{fig:ESO_Sky_Spectrum_Nitrogen}. }
 \label{fig:ESO_Sky_Spectrum_Oxygen}
\end{figure*}

In the WIM, oxygen atoms are excited by electron collisions to the
$^1{\rm D}_2$ level.  This level decays to the ground state ($^3{\rm
P}_2$) by emitting a [OI]\,6300 photon or to the first fine-structure
state ($^3{\rm P}_1$) by emitting a [OI]\,6364 photon.  The
A-coefficient branching ratio of three favors the former over the
latter.  An atom in the $^3{\rm P}_1$  state decays on a time scale
of $A_{ul}^{-1}=3.1\,$ h (by emitting a 63.18\,$\mu$m photon). The
resulting critical density, at $T\approx 10^4\,$K, is $3\times
10^5\,{\rm cm^{-3}}$. So, for the WIM, we can safely assume that
all oxygen atoms are at the ground level.

Using the collision strength from Appendix~F of \cite{D11} we find
 \begin{equation}
	\mathcal{I}_{6300}=6.5\Big(\frac{x_{\rm O^0}}{x_{\rm H^+}}\Big)
	\frac{T_4^{0.93}}{1+0.605T_4^{1.105}} {\rm EM}\,{\rm
	e}^{-2.283/T_4}\, R
  \label{eq:I_6300}
 \end{equation}
where we have assumed the abundance of oxygen, relative to hydrogen
by number, of 537\,ppm (Chapter 1 of \citealt{D11}).  At $T=10^4\,$K,
for the same EM, the intensity from [OI]\,6300 is nearly six times
stronger than that of [NI]\,5200 (see Equation~\ref{eq:I5200}).

The deepest observations of [OI]\,6300 from the WIM are from
\cite{rht+98}. WHAM observations towards a number of lines-of-sight,
all devoid of H~II regions, but along or close to the Galactic
equator, were undertaken. The authors reported that the photon
intensity ratios $\mathcal{I}_{6300}/\mathcal{I}_{\rm H\alpha}$,
are $<0.01$, 0.019, 0.027, 0.04, some two to seven times smaller
than computed by \citet{M86}.  Parenthetically, we note that the
$\mathcal{I}_{6300}/\mathcal{I}_{\rm H\alpha}$ ratio for H~II regions
is an order of magnitude weaker, ranging from $10^{-3}$ to $5\times
10^{-3}$ \citep{hrf02}.

\subsection{[OI] Airglow}
 \label{sec:OI_Airglow}

[OI] emission at 6300\,\AA{} and 6364\,\AA{} in the nocturnal
ionosphere of the Earth is produced in a way similar to that of the
[NI] airglow (see \S\ref{sec:Airglow_NI}; also, for
example,\citealt{kelley09}).  Collisions between the dominant ion
of the ionospheric F layer, ${\rm O}^{+}$ with ${\rm O}_2$ lead to
the formation of ${\rm O}^{+}_2$ ions. Subsequent collisions with
electrons can then produce excited O atoms in the crucial $^1{\rm
D}_2$ state by dissociative recombination. Excited O atoms are also
produced by direct recombination of ${\rm O}^{+}$ ions and electrons
\citep{slanger04}, but this is a slow process that can be neglected
for lines based on $^1{\rm D}_2$. As can be seen in
Figure~\ref{fig:ESO_Sky_Spectrum_Oxygen}, the airglow of [OI],6300
is strong.

\begin{figure}[htbp]   
 \plotone{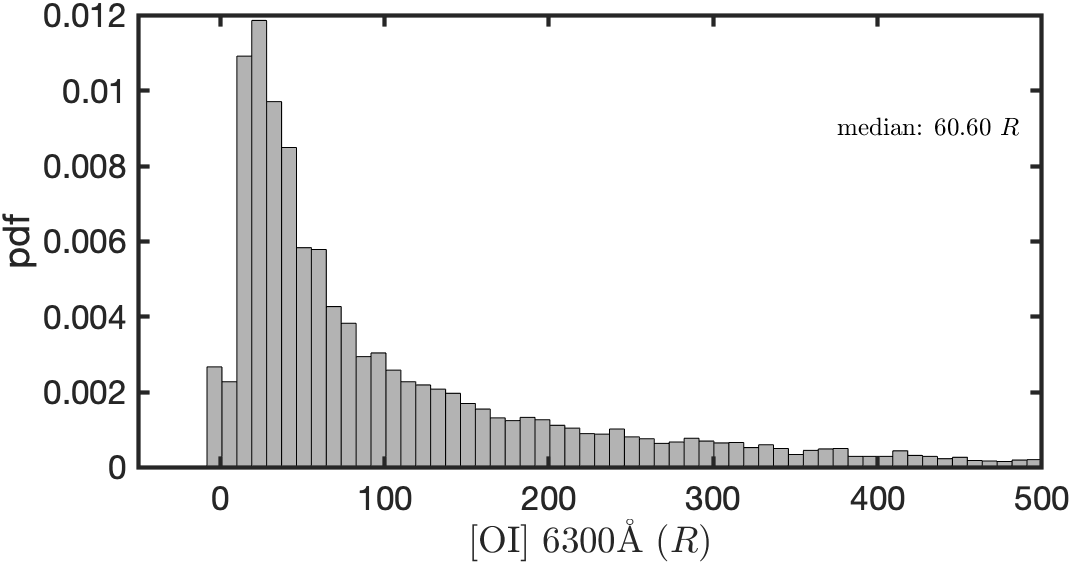}
  \caption{\small Histogram of [OI]\,6300 airglow emission (extrapolated
  to the zenith). }
 \label{fig:OI6300_histogram}
\end{figure}

In a manner similar to that discussed for the [NI] airglow
(\S\ref{sec:Airglow_NI}) we measured the intensities of the two [OI]
airglow lines in the X-shooter data (see also \S\ref{sec:X-shooter}).
The histogram of $\mathcal{I}_{6300}$ is shown in
Figure~\ref{fig:OI6300_histogram}.  The median airglow [OI]\,6300
is hundred times stronger than that of [NI]\,5200.

\begin{figure}  
 \plotone{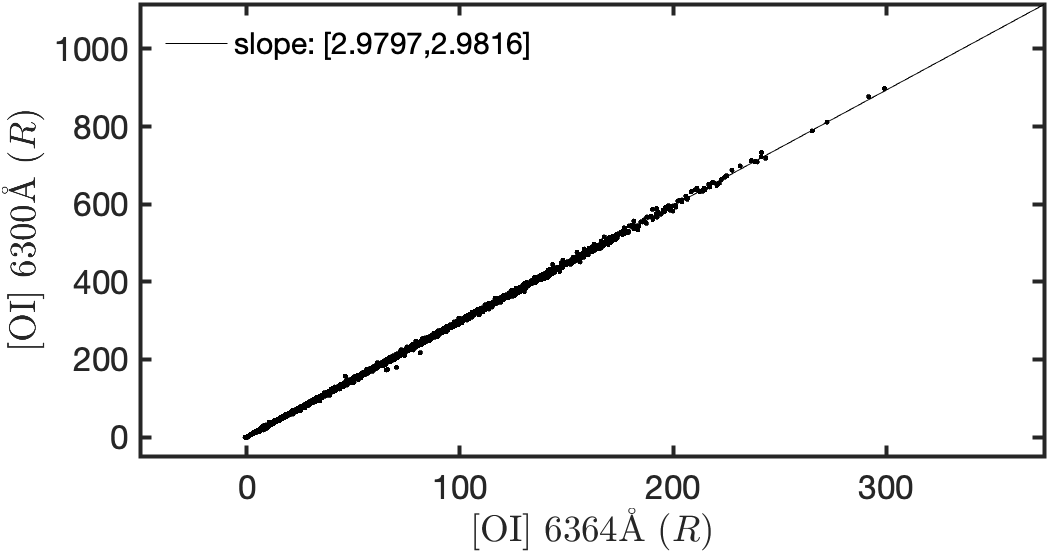}
  \caption{\small The correlation between the two nebular lines of
  O~I.  The 95\% confidence interval for the slope of the best fit
  line is shown in the figure.}
 \label{fig:OI_ratio}
\end{figure}

Next, we find that the [OI]\,6364 line faithfully tracks the
[OI]\,6300 with a slope of [2.980,2.982] (95\% confidence level;
see Figure~\ref{fig:OI_ratio}). Our determination is in good agreement
with the slope value of $2.997\pm 0.016$, deduced from Keck ESI and
HIRES measurements \citep{ss06}.  The two lines result from the
decay of a single excited state (${\rm ^1D_2}$) to levels in the ground
term, ${\rm ^3P}$. As a result, the relative photon intensities are
given by the branching ratio computed from the A coefficients (via both
E2 and M1). From NIST we find the branching ratio to be 3.099.  We
note that the accuracy noted by NIST for the A coefficients
is ``B+".  We parenthetically note that the WIM [OI]\,6364 has an
intensity comparable to that of [NI]\,5200 but an airglow background
that is about thirty times larger.

\section{Detectability}
 \label{sec:Detectability}

WHAM pioneered the study of the WIM in optical lines. However, only
a few lines-of-sight were observed in [OI]\,6300\citep{rht+98}.
{\it Only one line-of-sight was observed in [NI]\,5200} \citep{rrs77}.
A new generation of wide-field spectrographs have come up, and so
an investigation of detectability of WIM lines with these spectrographs
is timely.

Fabry-P\'erot spectrometers were popular in the nineties (e.g.,
WHAM).  The fashion has now shifted to wide-field IFU spectrographs.
In Table~\ref{tab:IFUs} we list some of the leading IFUs.  An optical
imaging instrument has two angular scales: the field-of-view
($\Omega$) and the size of pixel.  In contrast, for a simple radio
telescope, the FoV is composed of a single pixel (``beam").  The
same is the case for WHAM.  Our main focus is the study of the WIM
and for our purpose it is sufficient to use IFUs in a ``light-bucket"
mode. In this mode, the three-dimensional IFU data cube (two
orthogonal angular coordinates and a wavelength axis) is reduced
to a 1-D spectrum, obtained by replacing each channel image with
the sky value, usually the median.  This light-bucket mode maximizes
the sensitivity of the IFU for low surface-brightness studies.

\begin{deluxetable}{lrrrlr}[ht]
 \tablecaption{WHAM and IFU spectrographs}
\label{tab:IFUs}
\tablewidth{0pt}
\tablehead{
 \colhead{instrument} & 
 \colhead{$D$} & 
 \colhead{$\Omega_a$} &
 \colhead{$R_\lambda$} & 
 \colhead{$\eta$} & 
 \colhead{$\mathcal{G}$}\\
 \colhead{} & \colhead{(m)} & \colhead{(arcmin$^2$)}  & & &
 (cm$^2$\,arcmin$^2$) }
\startdata
LVM/SDSS$^a$  & 0.16  & 490 & 4000    & 0.25    & $0.98\times 10^8$ \\
KCI/Keck$^b$     & 10.0   & 0.0467 & 18000 & 0.25  & $1.65\times 10^8$\\
MUSE/VLT$^c$  & 8.20  & 0.983 & 2500    & 0.13    & $1.69\times 10^8$\\
\hline
WHAM$^d$ & 0.60 & 2827 & 30000 & 0.12 & $2.9\times 10^{10}$\\
\enddata
 \tablecomments{ $D$ is the diameter of the telescope, $\Omega_a$
 is the Field-of-View (FoV) in arcmin$^2$, $R_\lambda$ is the
 spectral resolution (defined in  \S\ref{sec:Sensitivity}) and
 $\eta$ is the photon-to-photoelectron efficiency.  $\mathcal{G}$
 is the ``spectral grasp" and is defined in \S\ref{sec:Sensitivity}.
 $^a$see \S\ref{sec:LVM} for a summary.  $^b$The Keck Cosmic Imager
 (KCI) consists of the blue spectrograph \citep{mmm+18} and the
 recently commissioned red spectrograph \citep{mmn+24}.  The
 instrument has many modes (trading between FoV, spectral resolution,
 and instantaneous wavelength coverage).  The highest spectral
 resolving power ($R_\lambda \approx 18,000$) results in the smallest
 FoV ($8.4^{\prime\prime}\times 20.4^{\prime\prime}$) and restricts
 the instantaneous wavelength interval to about 100\,\AA.
 $^c$\cite{baa+10}.  $^d$\citealt{T97,rth+98}.  The throughput for
 WHAM is from R.\ J.\ Reynolds (pers.\ comm.).  }
\end{deluxetable}

\subsection{Sensitivity}
	\label{sec:Sensitivity}
	
Consider a spectrometer with channel width,
$\Delta\lambda=\lambda/R_\lambda$ where $R_\lambda$ is the spectral
resolving power.  We assume that the signal is a line contained in
a single spectral channel.  It can severely underfill the channel
(low-resolution spectrographs).  $\mathcal{S}$, in Rayleigh, is the
signal integrated over the channel.  Let $\mathcal{B}$ be the sky
brightness carrying the unit Rayleigh per \AA.  We will assume that
the dominant noise source is the Poisson fluctuations induced by
the sky.

The number of signal photons in a given channel is
$\eta\alpha\mathcal{S}A\Omega t$ while that due to the background
is $\eta\alpha\mathcal{B}A\Omega t \lambda/R_\lambda$. Here, $A$
is the collecting area ($\pi/4 D^2$) in  cm$^2$, $\Omega$ is the
field-of-view (FoV) in steradians, $t$ is the integration time in
seconds, $\eta$ is the photon-to-photoelectron conversion efficiency
and $\alpha=10^6/(4\pi)$ is the conversion factor between the intensity
in Rayleigh to  the photon intensity in the CGS system (${\rm
phot\,cm^{-2}\,s^{-1}\,ster^{-1}}$).  The signal-to-noise ratio
(SNR) of any channel is
 \begin{equation*}
  {\rm SNR} = (\eta\alpha A\Omega t)^{1/2}\frac{\mathcal{S}}
  {\sqrt{\mathcal{S}+\mathcal{B}\lambda/R_\lambda}}
 \end{equation*}

It is more convenient to specify the field-of-view in units of
arcmin$^2$, $\Omega_a$, instead of steradians.  We also simplify
by considering the case where the background dominates,
$\mathcal{B}\Delta\lambda\gg\mathcal{S}$.  The resulting SNR is
 \begin{equation*}
	{\rm SNR} \approx 636
	\Big(\frac{\lambda}{6,000\,\textup{\AA}}\Big)^{1/2}\frac{\mathcal{S}}{\sqrt{\mathcal{B}}}
	\Big(\frac{\mathcal{G}}{10^8\,{\rm
	cm^2\,arcmin^2}}\Big)^{1/2}t_{\rm hr}^{1/2}\ .
 \end{equation*}
where $\mathcal{G}= \eta A\Omega_{a}R_\lambda\, {\rm cm^2\,arcmin^2}$
is the spectral grasp and $t_{\rm hr}=t/{\rm 1\,hour}$.

The continuum background for both lines is roughly $1\,R\,$\AA$^{-1}$.
So, for simplicity, we set $\mathcal{B}=1\,R\,$\AA$^{-1}$ below.
The desired signal levels are summarized in Figure~\ref{fig:OI_NI}.
We adopt $T_*=36,000\,$K.  For [OI], the expected $\lambda$6300/H$\alpha$
is $9.7\times 10^{-3}$ for ${\rm log}(u)=-3$ and $3.0\times 10^{-2}$
for ${\rm log}(u)=-4$.  For [NI], $\lambda$5200/H$\alpha$ is
$1.7\times 10^{-3}$ and $6.3\times 10^{-3}$, respectively.

Of the IFUs listed in Table~\ref{tab:IFUs} only KCI has spectral
resolution to marginally resolve the airglow emission from the WIM
emission.  Thus, KCI is well suited for both [OI] and [NI] observations.
As discussed in \S\ref{sec:NI_DoubletMethod}, [NI] can be usefully
observed by any spectrometer that can resolve the [NI] V-band
2.5-\AA\ doublet. MUSE seems marginally suitable.  LVM resolves the
[NI] V-band doublet (N.\ P.\ Konidaris, pers.\ comm.).  In the
following, we investigate WIM studies with KCI and LVM.


\subsection{Keck Cosmic Imager}
	\label{sec:KCI}
	
We will assume that KCI will be used with the highest spectral
resolution so that the airglow and WIM contributions are separated
in the spectrum.  The rms noise per channel for KCI is $0.0012t_{\rm
hr}^{-1/2}\,R$.   Thus, at mid-latitudes, it should be possible to reach
$\lambda$6300/H$\alpha$ of $1/300$ with, say, a few hours of integration
time.  This level of sensitivity is sufficient for [OI] measurements
to be very useful.  Observations can detect [NI]\,5200 only if low
values of $u$ hold.

\subsection{The Local Volume Mapper}
 \label{sec:LVM}
 
The Local Volume Mapper (LVM; \citealt{dbk+24}) is one of the three
``mappers" of the Sloan Digital Sky Survey Phase V (SDSS-V;
\citealt{kzr+17}).  The facility consists of four 16-cm telescopes:
Science, Sky1, Sky2 and Spectrophotometry with 1801, 60, 59
and 24 fibers at the focal plane \citep{bmb+24,fzk+20,hbb+24} which
are fed to a bank of spectrographs \citep{kdf+20}. Each fiber
subtends a beam of diameter 35.3$^{\prime\prime}$ on the sky. The
Sky telescopes are mainly devoted to measuring geocoronal
H$\alpha$. The Sky telescopes flank the Science telescope
and are directed to regions of low emission measure.  The solid
angle covered by the Science telescope is about 490\,arcmin$^2$
or 0.136\,deg$^2$.  The spectrographs cover the spectral range
3600--9800\,\AA\ with a spectral resolution of about 4,000.  With
15 minutes of integration, H$\alpha$ is expected to be detected
by the {\it science} telescope at an intensity of $6\times
10^{-18}\,{\rm erg\,cm^{-2}\,s^{-1}\,arcsec^{-2}}$, which corresponds
to 1.06,$R$ at a signal-to-noise ratio (SNR) of 5.

\begin{figure}[htbp]     
 \plotone{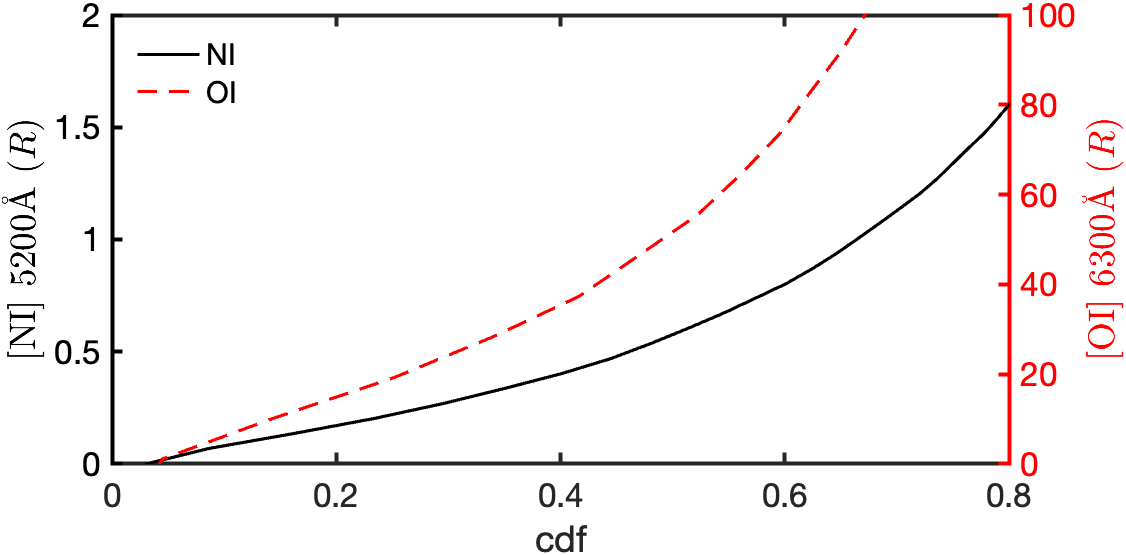}
  \caption{\small The cumulative probability distribution function
  (cdf; $x$-axis)  of [NI]\,5200 (left)  and [OI]\,6300 (right)
  airglow intensities. The two lines are highly correlated.  
  Note the differing intensity scales for [OI]
  relative to [NI].  }
 \label{fig:NI_OI_cdf}
\end{figure}

As noted above, we can use LVM to observe [NI]\,5200 with the doublet
method (\S\ref{sec:NI_DoubletMethod}) but with additional loss of
performance owing to the penalty resulting from the extraction of
the WIM component from the airglow line.  As can be seen in
Figure~\ref{fig:NI_OI_cdf} the median intensity of the [NI]\,5200
line\footnote{we account for the fact that  airglow [NI]\,5198 is
larger than [NI]\,5200 by a factor of 1.74.} is well below the sky
continuum ($\approx 1\,R\,\textup{\AA}^{-1}$).  Thus, the performance
of LVM for [NI]\,$\lambda\lambda$\,5198,\,5200 will, in  most instances,
be limited by the sky continuum and not the [NI] airglow emission.

\section{Summary \&\ Conclusions}
 \label{sec:SummaryConclusions}

The WIM is a major phase of the Galactic ISM. The primary data --
recombination lines (primarily H$\alpha$) and optical nebular
lines of N$^+$ and S$^+$ -- are adequately explained by photoionization
models with low ionization parameter, $u\approx 10^{-3}$ to $10^{-4}$.
H~II regions have $u\gtrsim 10^{-2}$.  Relative to H~II regions,
low-ionization nebulae have two distinct characteristics: ({\it i}) the
transition from the ionized interior to the neutral medium is not
sharp and ({\it ii}) the mean neutral fraction can range from a few
percent to tens of percent.  In partially ionized plasma, charge
exchange can become a major player\footnote{For instance, the charge
transfer coefficient for the reaction ${\rm N^{+2} + H^0 \rightarrow
N^+ + H^+}$ is three orders of magnitude larger than the inverse
reaction and two orders of magnitude larger than the radiative
recombination rate, ${\rm N^{+2}+e^- \rightarrow N^+}$.  As a result,
N$^{+2}$ will rapidly convert to N$^+$.  This explains why
nitrogen in the WIM is primarily in the form of N$^+$.  The same
applies to O$^{+2}$.  However, charge exchange does not favor
Ne$^{+2}$ whereas ${\rm H^++Fe^+\rightarrow H+Fe^{+2}}$ explains the
large abundance of Fe$^{+2}$ in the planetary nebulae \citep{D85}.},
leading to the suppression of highly charged ions.  The transition
region is warm enough to still excite neutral species and the
dominant lines are [OI]\,$\lambda\lambda\,6300,\,6363$ and
[NI]\,$\lambda\lambda$ 5198,\,5200.  Thus, these lines serve as
excellent probes of the transition region.

An important topic in Galactic ISM research is the ionization and
heating of the gas in the halo of the Galaxy (``interstellar halo").
\cite{dm94} noted that the [OI]\,6300 line is a direct measure of
the fraction of the diffuse Lyman continuum absorbed by the WNM.
So, in addition to being a diagnostic of the transition region,
[OI]\,6300 can potentially be used to infer the escape fraction of
the diffuse Lyman continuum into the interstellar halo.

This paper is focused on the study of neutral gas within the WIM.
Our current understanding of the neutral gas within the WIM is
unsatisfactory.  As can be gathered from Figure~\ref{fig:OI_NI}
there is degeneracy between $u$ and $X_{\rm edge}$ for the amount
of neutral gas within the WIM. With little doubt a sizable fraction
of Lyman continuum rays will be absorbed by WNM ($X_{\rm  edge}=0.98$).
The others may escape directly to the halo or encounter thin slabs.
The path forward is to build solid statistics, which means
observations neutral lines towards many tens of
lines-of-sight.  Unfortunately, as noted several times in this
paper, observations in [OI] and [NI] are lacking.

Traditionally, the [OI]\,6300 line has been used for
this purpose.  However, it turns out that [OI]\,6300 is a bright
line, so one needs a high spectral resolution to separate the WIM
and airglow contributions. Most modern IFUs lack the exquisite
spectral resolution required to study [OI].

In this paper, we explored the feasibility of observing the
nebular lines of [NI].  The ionization potential of nitrogen is
only about 1\,eV higher than that of hydrogen, so N$^0$ should
be a good tracer of warm H$^0$.  The [NI]\,5200 line, owing to its
smaller abundance, is six times weaker than the [OI]\,6300 line.
However, the [NI]\,5200 airglow line is a hundred times weaker
than the [OI] airglow line.  Furthermore, it is not necessary to
resolve the airglow line from the WIM. We show that the observed
doublet ratio of [NI] provides a good measure of the WIM intensity
provided that the airglow line is comparable to or weaker than the WIM
component.  We find that with suitable planning (see below)
observations can be undertaken during periods in which airglow [NI]
emission is weak.  Furthermore, for this purpose, wide-field
spectrographs with modest resolution, $\gtrsim 4,000$, will suffice.

During the past decade, several powerful wide-field IFUs have been
commissioned (Table~\ref{tab:IFUs}).  Parenthetically, we note that
they have comparable spectral grasp.  The Keck Cosmic Imager and
the Local Volume Mapper have the capability to carry out [NI] studies
of the WIM.  In addition, KCI, with some planning, has adequate
resolution to undertake [OI] studies of the WIM.

R.\ Reynolds (1943--2024) pioneered  the study of the WIM by
introducing large etendue Fabry-P\'erot  spectrometers, culminating
with WHAM.  Wide-field studies of the Galactic sky in H$\alpha$,
[SII], and [NII] are his lasting legacy.  As can be seen in
Table~\ref{tab:IFUs}, WHAM, despite being commissioned more than
three decades ago, remains a unique instrument.  Here, we offer
compelling reasons for a focused [OI]\,6300 program with WHAM. 

Separately, we note that an additional escape mechanism becomes
possible if the filling factor of the HIM is comparable to that of
the WNM.  In this case, an ionizing ray could reach the halo,
unhindered by WNM.  This mechanism certainly operates in certain
regions of the Galaxy (e.g., Galactic chimneys).  This escape
fraction is obviously not captured by [OI] or [NI] tracers. An
elegant way\footnote{The primary complication is possible ionizing
radiation arising from the HIM-cloud interface.}  to measure the
{\it total} escape fraction is by measuring the H$\alpha$ fluorescence
of high velocity clouds with known distances (e.g.  \citealt{rtk+95}).
In short, there is a pressing need for WHAM and LVM to undertake a
systematic program of deep observations of intermediate- and
high-velocity clouds. LVM has the added value of spatially resolving
the structure of the fluorescent surfaces of HVCs.

\subsection{Dynamic Scheduling}

As discussed in \S\ref{sec:Optimizing} the airglow emission is
highly variable.  The dynamic night sky makes it important for
astronomical observatories to have their own ``airglow" monitor.
There have been major airglow programs at Paranal.  The effort
includes analysis of archived data (discussed at various points in
this paper) as well as all-sky imaging at high temporal
resolution\footnote{\url{https://www.eso.org/public/teles-instr/paranal-observatory/oasis/}}.

For our purpose a simple airglow spectrum monitor would suffice.
This facility does not require significant investment.  Inexpensive
commercial wideband fiber-fed spectrometers, with good spectral
resolution, are now available\footnote{e.g., \url{www.jinsp.com},
\url{www.avantes.com}}.  For the telescope, a simple 6-inch lens
would suffice.  The resulting real-time sky spectrum would inform
astronomers of the airglow situation (along the line-of-sight towards
a planned target).  Astronomers could dynamically schedule the
observations to observe their target during periods of low airglow
activity.

Separately, [OI] (and also [NI]) airglow emissions are
particularly intense and variable in two bands, approximately $15^\circ$
above and below the magnetic equator.  These are the crests of the
``equatorial ionization anomaly" (see, e.g., \citealt{esd+19}).
They are related to increased ion and electron densities in the
ionospheric F layer. These structures are readily observed by airglow
satellite missions.  
In this spirit, we suggest that astronomers invite airglow
researchers for future siting of the next generation of WHAM.


\subsection{Diffuse Ionized Gas}


Although our focus has been on the Galactic WIM, some of the conclusions
in this paper are applicable to the study of the DIG (in external
galaxies).  To start with, extragalactic studies are not plagued
by the bright [OI]\,6300 airglow.  We discuss two examples of the
potential value of [NI] observations.  NGC 5291N is a dwarf galaxy
undergoing copious star-formation. \citet{fdw+16} present deep MUSE
observations of NGC 5291N.  Strong [OI] is detected in the DIG. It
appears to us that [NI]\,5200 was also detected, although not
commented upon by the authors.  MUSE observations of the Antennae
galaxy system show strong [OI]\,6300 in the outer regions
\citep{wmv+18,gbw+20}.  Given the strong detection of [OI] we expect
[NI]\,5200 should also be present in the data.  For both of these cases,
we point out that $\lambda$6300/$\lambda$5200 provides an independent
O/N ratio that is relatively immune to temperature, extinction, and
ionization state.

\subsection{A bright future}

We end this paper by noting that the study of the WIM, thanks to
new instruments, has a bright future.  To date, the abundance of
metals in the WIM have come primarily from optical studies with
WHAM, primarily via [SII] and [NII] observations.  WHAM did not
have good blue sensitivity. As a result, WHAM did not observe in
[OII]. This is unfortunate since oxygen is not only the most abundant
metal in the WIM but is also the strongest coolant.  Fortunately,
there are new beginnings. \cite{mrr+06} undertook the first WIM
observations in [OII], using a new method (spatial heterodyne
spectrometer). Separately, LVM has the sensitivity to detect [OII].

Next, MIRI-MRS, a formidable IFU aboard the James Webb Space Telescope
(JWST), is now working routinely. \cite{kbr24} showed that the WIM
can be routinely detected, via fine-structure lines of Ne$^+$,
S$^{+2}$ and Ar$^+$, in ``off-beam" data. The ever increasing
archival data will, in due course, yield abundances of two noble
gases in the WIM and also of sulfur, when combined with optical
[SII] observations.

The primary limitation with WHAM has been the one-degree beam. This
is unfortunate since the WIM, like the WNM and CNM, is bound to
have considerable angular structure.  Fortunately, LVM has the
capacity to study the WIM on arc-minute scales.  A few dark weeks
set aside for deep high-latitude studies of the WIM would be novel
and likely revolutionary.

More than six decades ago, the WIM was identified as a pervasive
medium from low-frequency radio observations \citep{ewb62,he63}.
Modern low frequency facilities such as
LOFAR\footnote{\url{https://www.astron.nl/telescopes/lofar/}},
MWA\footnote{\url{https://www.mwatelescope.org/telescope/}} and
OVRO-LWA\footnote{url{https://www.ovro.caltech.edu/projects}} have
now come of age and should soon start providing detailed information
on the WIM but on arc minute scales.

\acknowledgements

SRK thanks  (the late) Ron Reynolds for several discussions, during
the covid period, about the WIM and WHAM.  We thank the following
for their feedback: Pieter van Dokkum, Jason Spyromilio and Chuck
Steidel.  Stefan Noll was financed by the project NO 1328/1-3 of
the German Research Foundation (DFG).

 \bibliography{bibNitrogen}{}
\bibliographystyle{aasjournal}

\appendix

\section{Simple 1-D modeling}
 \label{sec:1-D}
 
The WNM slab (hydrogen nuclei density, $n_{\rm H}$) occupies the
volume $z\ge 0$ with front face defined by $z=0$ upon which a diffuse
Lyman continuum, photon intensity, $I_0$, shines. By construction,
the slab is thick enough that there is no intensity coming from the
``under side". The density of ionizing photons at $z=0$ is $n_\gamma=\pi
I_0/c$ and the corresponding ionization parameter is $u=n_\gamma/n_{\rm
H}$.

We make the ``case B" approximation: free-bound photons resulting
from recombinations to the $n=1$ level of hydrogen are absorbed
close to the site of the emission and photons resulting from emission
to any other level will escape from the nebula.  With this assumption,
the primary ionizing source is the incident diffuse Lyman continuum.
Thus,
 \begin{equation*}
	\frac{dI(z)}{dz} = - \sigma_{\rm pi}n_{\rm H^0}I(z)
\end{equation*} where $\sigma_{\rm pi}$ is the photoionization
cross section.  This equation can be recast as
 \begin{eqnarray}
  \frac{d\mathcal{I}}{d\tau} &=& -(1-x)\mathcal{I}
	\label{eq:radiative_transfer}
 \end{eqnarray}
where $\mathcal{I}(z)=I(z)/I_0$, $x=n_e/n_{\rm H}$ is the ionization
fraction, $n_e$ ($n_{\rm H^+}$) is the number density of electrons
(protons) and $d\tau=n_{\rm H}\sigma_{\rm pi}dz$.  The local
volumetric ionization rate must be balanced by the local volumetric
recombination rate, which means $\pi I\sigma_{\rm pi} n_{\rm H^0}
= \alpha n_e^2\ $ where $\alpha$ is the recombination coefficient.
This equality can be recast as
 \begin{equation}
	(1 -x)\mathcal{I}= \frac{x^2}{\xi}
		\label{eq:ionization_balance}
 \end{equation}
where $\xi= \pi I_0\sigma_{\rm pi}/(n_{\rm H}\alpha)$.  Note that
$\xi$, a dimensional-less parameter, is simply the ratio of the
recombination timescale to the photon-ionization timescale.  The
case B recombination coefficient at $T=8,000\,$K is $\alpha=3\times
10^{-13}\,{\rm cm^3\,s^{-1}}$.  For a typical ionizing photon energy
of 20\,eV, we have $\sigma_{\rm pi}= 2\times 10^{-18}\,{\rm cm^2}$.
For these values, $\xi\approx 2\times 10^5u$ where $u=n_\gamma/ n_{\rm
H}$.  Combining Equations~\ref{eq:radiative_transfer} and
\ref{eq:ionization_balance} yields $d\mathcal{I}/d\tau = -x^2/\xi$
whose integration results in
 \begin{equation*}
	1- \mathcal{I}(\tau^\prime) = \int_0^{\tau^\prime} x^2
	d\tau^\prime
 \end{equation*}
where $\tau^\prime=\tau/\xi$ is the ``scaled" optical depth
and is the natural primary variable for this model.  The ionizing
intensity will, as it progresses into the slab, be attenuated and
asymptotically approach zero. Thus, $\int_0^\infty x^2d\tau^\prime=1$,
which, upon substitution for $d\tau^\prime$, leads to $\pi I_0 =
\int_0^\infty \alpha n_e^2 dz$ -- that is, the rate of recombinations
in the entire column with a base of of 1\,cm$^2$  is equal to the
flux of ionizing photons entering the slab.

Equations~\ref{eq:radiative_transfer} (a differential equation) and
\ref{eq:ionization_balance} (a quadratic equation in $x$) are the
governing equations.  To solve Equation~\ref{eq:radiative_transfer}
we choose a small step, $\Delta\tau$.  We start by noting that
$\mathcal{I}(\tau)=1$ at $\tau=0$.  We substitute this value of
$\mathcal{I}$  into Equation~\ref{eq:ionization_balance} and solve
for $x$.  Armed with a value for $x$, we use
Equation~\ref{eq:radiative_transfer} to compute $\mathcal{I}(\Delta\tau)$.
We consider the edge of the WIM to be marked by $x=0.1$ (depth,
$L$).  The corresponding total optical depth is $\tau_0=n_{\rm H}\sigma_{\rm
pi}L$.

\begin{deluxetable}{lrrrrrrr}[hbt]  
\tabletypesize{\footnotesize}
\tablewidth{0pt}
\tablecaption{A slab ionized by diffuse  continuum}
\label{tab:xXg_slab}
\tablehead{
\colhead{$\xi$} &
 \colhead{${\rm log}(u)$} &
 \colhead{} &
 \colhead{$X(0)$} &
\colhead{$\langle X_{90}\rangle$} &
\colhead{$\tau_e^\prime$}  &
\colhead{$g$} &
\colhead{$\tau_0^\prime$}
}
\startdata
10 &  $-4.30$ &  & 0.084  & 0.34 & 0.75   & 1.32 & 2.00\\ 
30 &  $-3.82$ &  & 0.031  & 0.19 & 0.87   & 1.14 & 1.40\\ 
100 &  $-3.30$ &  & 0.010  & 0.08 & 0.95   & 1.05 & 1.15\\ 
300 &  $-2.82$ &  & 0.003  & 0.03 & 0.98   & 1.02 & 1.06\\ 
\enddata
\end{deluxetable}

We summarize useful physical parameters in Table~\ref{tab:xXg_slab}.
$X(0)$ is the neutral fraction at $z=0$, while the neutral fraction
at the bottom of the WIM layer is 90\% (\S\ref{sec:SimplifiedModel}).
$\langle X_{90}\rangle$ is the mean neutral fraction in the range
$X=[X(0),90\%]$.   The dispersion measure, $\int n_edl\propto
\tau_e\equiv \int_0^{\tau_0}xd\tau$.  Finally, $g$ is the ratio of
the emission measure to the dispersion measure, $g=\int n_e^2dl/\int
n_edl$, $\tau_e^\prime=\tau_e/\xi$ and $\tau_0^\prime=\tau_0/\xi$.

\section{X-shooter Observations}
\label{sec:X-shooter}

The X-shooter is currently mounted on the Melipal telescope of the
European Southern Observatory, Cerro Paranal, Chile (24.6$^{\circ}$\,S,
70.4$^{\circ}$\,W). It is a medium resolution wideband multi-armed
echelle spectrograph \citep{vdd+11}. Archival data of the three
arms: UVB (3000 to 5600\,\AA), VIS (5500 to 1,0200\,\AA), and NIR
(1.02\,$\mu$m to 2.48\,$\mu$m) are well suited to study terrestrial
airglow. With the projected length of the entrance slit of
$11^{\prime\prime}$, it is possible to separate the airglow and the
astronomical target, especially if the latter is point-like and
faint. Therefore, X-shooter spectra have already been used to study
the chemiluminescent emissions of hydroxyl, molecular oxygen, sodium,
iron monoxide, and hydroperoxyl
\citep{noll15,noll16,unterguggenberger17,noll22,noll23,noll24}.
The astronomical observations are performed with different set-ups
(slit width and binning) and exposure times differing by several
orders of magnitude. As a result, the successful use of these spectra
for investigation of airglow phenomena requires a thorough data
selection.

Our primary interest are [NI] and [OI] lines, which require only
UVB- and VIS-arm spectra, respectively.  Our data set of spectra was
obtained between October 2009 (the beginning of the archive) and September
2019. This data set was already prepared for other airglow-related
studies. The details of the production of one-dimensional,
wavelength-calibrated, and flux-calibrated sky spectra are described
by \citet{noll22}; for the NIR dataset, see \citet{noll23}. 
Data for all arms are presented by \citet{noll24}. Due
to a thorough derivation of instrumental response curves, the
uncertainties in the flux calibration should only be of the order
of a few percent for clear sky conditions.

Line measurements were performed in an automated manner (see also
\citealt{noll22}). The underlying continuum was estimated using a
wide median filter for [NI]. For [OI], due to the increase in the
density of lines in the red wavelength range, a 45-percentile filter
was employed.  In both cases, the filter width was 0.008 times the
wavelength.  Line integrations at the expected wavelengths (drawn
from the NIST portal) in the continuum-subtracted spectra were
undertaken with integration ranges depending on the slit width.
Unusual continua or uncertainties in the wavelength calibration
(normally less than 1 pixel) could cause significant changes in the
intensity.

The measured line intensities were corrected for different effects.
First, we corrected for atmospheric absorption and scattering.  For
the measured lines, the absorption is essentially caused by ozone
and amounts to about 1\% at 5200\,\AA\  and about 3\% at 6300\,\AA\
at zenith and increases towards the horizon for the absorption
spectra of the Cerro Paranal sky model \citep{nkb+12}.  Natural
changes in the column density of ozone were not considered because
these changes cause only very minor uncertainties in line intensities.
The changes resulting from the scattering of airglow photons at
atmospheric molecules (Rayleigh scattering) and aerosol particles
(Mie scattering) were also corrected by means of the recipes of
\citet{nkb+12}. At zenith, this results in a slight increase (plus
2.4\% at 5200\,\AA\ and 1.4\% at 6300\,\AA), whereas there is a
clear decrease at a zenith angle of 60$^{\circ}$ (minus 6.7\% and
3.3\%, respectively). The uncertainties of these corrections are
relatively high (but still small with respect to the line intensity),
since the aerosol composition and densities are not well known
\citep[see][]{jones19}, and the corrections were derived for an
emission altitude of only 90\,km. Finally, intensities tend to
increase with increasing zenith angle due to the larger apparent
width of the emission layer in the viewing direction \citep{vanrhijn21}.
We corrected for this increase by assuming thin layers with reference
heights of 200\,km for the [NI] lines \citep{ssc+05} and 250\,km
for the red [OI] lines (slightly below the peak of the F2 layer).
The correction factors only weakly depend on the reference heights,
at least for the relatively low zenith angles that are typical of
astronomical observations.

As already mentioned, a thorough sample selection is important for
a reliable data set. For this purpose, we checked the quality of
the line measurements based on different observing parameters.
As a consequence, we rejected all exposures of the UVB arm, which
includes the weak [NI] doublet, with durations of less than 10 minutes.
We also excluded all observations which showed a strong underlying
continuum (resulting from scattered moonlight or extended astronomical
sources). In practice, we rejected data sets with a continuum
brightness of $> 1.5\,R\,$\AA$^{-1}$ in the vicinity of the [NI]
lines. In addition, spectra with unusually weak continuum ($<
0.35\,R\,$\AA$^{-1}$), possibly due to data processing issues, were
also excluded. In some cases (e.g. the Orion molecular complex)
the astronomical [NI] lines are extremely strong. For this reason,
all spectra with “OM” targets were rejected. Next, we note that the
[NI] doublet lines are only separated by 2.4\,\AA, which is too
close for the resolving power of observations with wide slits. For
this reason, we only included the slit widths $\le 1^{\prime\prime}$.
Finally, a comparison of the intensities of both doublet components
showed 10 clear outliers, which were also rejected. In the end, the
final sample amounts to 8,955 of the available 91,553 UVB-arm
spectra.

The above selection is also the basis for the red [OI] lines, which
are distinctly brighter (suggesting more robust measurements) but
are located in the X-shooter VIS arm.  The overhead times for the
UVB- and VIS-arm exposures appear to be different, which causes
small changes in the time coverage. Moreover, it sometimes happened
that several VIS-arm exposures were performed during a single UVB-arm
exposure. In this case, the resulting [OI] intensities represent a
mean weighted by the individual exposure times. On occasions, the
reduction pipeline appeared to have failed to produce valid data
products. As this can happen to just one X-shooter arm, there is
not always a VIS-arm for a UVB-arm spectrum. As a consequence, only
8,768 [OI] intensities can be compared to the [NI] intensities,
which corresponds to a loss of about 2\%, relative to the UVB-arm
data. For a reliable comparison, it also needs to be considered
that [OI] measurements can also suffer from significant systematic
effects, although the lines are much brighter and the continuum
tends to be less problematic.  Specifically [OI]\,6300 intensities
below $1\,R$ (30 cases) do not seem to be reliable.

\end{document}